\documentclass[journal]{IEEEtran}
\usepackage{amsmath,graphicx,amsfonts,multirow,cite}
\usepackage{todonotes}
\usepackage{tikz}
\usetikzlibrary{bayesnet,arrows,calc}

\newcommand{\D}{D}
\newcommand{\Q}{Q}

\newcommand{\W}{\mathbf{W}}
\newcommand{\F}{\mathbf{F}}

\newcommand{\V}{\mathbf{V}}
\newcommand{\Ss}{\mathbf{S}}
\newcommand{\fk}{\mathbf{f}_{k}}
\newcommand{\fo}{\mathbf{f}_{1}}
\newcommand{\fN}{\mathbf{f}_{N}}
\newcommand{\x}{\mathbf{x}}
\newcommand{\xh}{\hat{\mathbf{x}}}
\newcommand{\vv}{\mathbf{v}}
\newcommand{\h}{\mathbf{h}}
\newcommand{\s}{\mathbf{s}}
\newcommand{\sh}{\hat{\mathbf{s}}}
\newcommand{\shh}{\hat{s}}
\newcommand{\y}{\mathbf{Y}}
\newcommand{\z}{\mathbf{Z}}
\newcommand{\zh}{\hat{\mathbf{Z}}}
\newcommand{\m}{\mathbf{M}}
\newcommand{\zo}{\mathbf{Z}_{1}}
\newcommand{\zt}{\mathbf{Z}_{2}}
\newcommand{\zB}{\mathbf{Z}_{B}}
\newcommand{\zb}{\mathbf{Z}_{b}}
\newcommand{\yo}{\mathbf{Y}_{1}}
\newcommand{\yt}{\mathbf{Y}_{2}}
\newcommand{\yb}{\mathbf{Y}_{b}}
\newcommand{\yB}{\mathbf{Y}_{B}}
\newcommand{\Sop}{\mathcal{S}}
\newcommand{\Aop}{\mathcal{A}}
\newcommand{\No}{N_{o}}
\newcommand{\To}{T_{o}}
\newcommand{\Fs}{F_{s}}

\newcommand{\wz}{\omega_{0}}
\newcommand{\wo}{\omega_{1}}
\newcommand{\wt}{\omega_{2}}
\newcommand{\wB}{\omega_{B}}
\newcommand{\wBmo}{\omega_{B-1}}
\newcommand{\wbmo}{\omega_{b-1}}
\newcommand{\wb}{\omega_{b}}
\newcommand{\fenc}{\mathcal{E}}
\newcommand{\fdec}{\mathcal{D}}
\newcommand{\fmask}{\mathcal{M}}
\newcommand{\gmin}{G_{min}}
\newcommand{\preemp}{\beta}
\newcommand{\mupar}{\mu}
\newcommand{\fmu}{f_{\mu}}
\newcommand{\prior}{\pi}
\newcommand{\overcomp}{\kappa}

\title{A Multiscale Autoencoder (MSAE) Framework for End-to-End Neural Network Speech Enhancement}
\author{Bengt J. Borgstr$\ddot{\textrm{o}}$m and Michael S. Brandstein}

\begin{document}

\maketitle

\begin{abstract}
Neural network approaches to single-channel speech enhancement have received much recent attention. In particular, mask-based architectures have achieved significant performance improvements over conventional methods. This paper proposes a multiscale autoencoder (MSAE) for mask-based end-to-end neural network speech enhancement. The MSAE performs spectral decomposition of an input waveform within separate band-limited branches, each operating with a different rate and scale, to extract a sequence of multiscale embeddings. The proposed framework features intuitive parameterization of the autoencoder, including a flexible spectral band design based on the Constant-Q transform. Additionally, the MSAE is constructed entirely of differentiable operators, allowing it to be implemented within an end-to-end neural network, and be discriminatively trained. The MSAE draws motivation both from recent multiscale network topologies and from traditional multiresolution transforms in speech processing. Experimental results show the MSAE to provide clear performance benefits relative to conventional single-branch autoencoders. Additionally, the proposed framework is shown to outperform a variety of state-of-the-art enhancement systems, both in terms of objective speech quality metrics and automatic speech recognition accuracy. 
\end{abstract}


\section{Introduction}
\label{sec:introduction}
{\let\thefootnote\relax\footnote[0]{DISTRIBUTION STATEMENT A. Approved for public release. Distribution is unlimited.
This material is based upon work supported by the Department of Defense under Air Force Contract No. FA8702-15-D-0001. Any opinions, findings, conclusions or recommendations expressed in this material are those of the author(s) and do not necessarily reflect the views of the Department of Defense.
\par
$\copyright$ 2023 Massachusetts Institute of Technology.
\par
Delivered to the U.S. Government with Unlimited Rights, as defined in DFARS Part 252.227-7013 or 7014 (Feb 2014). Notwithstanding any copyright notice, U.S. Government rights in this work are defined by DFARS 252.227-7013 or DFARS 252.227-7014 as detailed above. Use of this work other than as specifically authorized by the U.S. Government may violate any copyrights that exist in this work.
}}
When captured in realistic acoustic environments, speech signals typically suffer from distortions such as additive noise and reverberation. For human listeners, this can lead to reduced intelligibility \cite{loizou2010} and increased listener fatigue \cite{zekveld2011}. It can also result in performance degradation for automated speech applications such as speech and speaker recognition \cite{borgstrom2012,virtanen2012,mandasari2012,li2014,sadjadi2014}. Speech enhancement can be employed to improve the perceptual quality of captured signals, and to curb these negative effects. For many decades, speech enhancement relied on statistical model-based techniques \cite{Mcaulay1980,Ephraim1984,Ephraim1985,Cohen2002}. Recently, however, deep neural networks (DNNs) have achieved impressive performance improvements due to their high modelling capacity \cite{Borgstrom2021,weninger2015,valentini2016,Zhao2018,Pandey2018,Macartney2018,Germain2018,Luo2018,Soni2018,Fu2018,Koizumi2018,bagchi2018,Pandey2019,Fu2019,Giri2019,Xu2019,Luo2019,Casebeer2020,Xu2020,Kolbaek2020,defossez2020,Li2021,fu2021,subakan2021,zhang2022}.
\par
Neural network speech enhancement systems can be categorized into \textit{generative} and \textit{mask-based} approaches. Generative systems are regression-based and synthesize the enhanced waveform using a set of learned filters, generally operating without much constraint on the output \cite{weninger2015,valentini2016,bagchi2018,defossez2020,fu2021,subakan2021,zhang2022}. While they can freely synthesize output waveforms and potentially reconstruct speech signal components that have been attenuated due to channel effects, they can also result in distorted speech signals or unnatural residual noise. Mask-based approaches, on the other hand, are constrained to apply a multiplicative mask in a time-frequency space in order to suppress interfering signal components \cite{Borgstrom2021,Zhao2018,Pandey2018,Macartney2018,Germain2018,Luo2018,Soni2018,Fu2018,Koizumi2018,Pandey2019,Fu2019,Giri2019,Xu2019,Luo2019,Casebeer2020,Xu2020,Kolbaek2020,Li2021}. While this constraint can help avoid distorted speech, it also limits the ability of the system to reconstruct any signals components not present in the original waveform.  
\par 
This paper focuses on mask-based approaches to neural network speech enhancement. In such systems, an encoder first transforms the input signal into an embedding space. A multiplicative mask is then applied to suppress signal components from interfering sources. Finally, a decoder synthesizes the output waveform. A crucial component of mask-based end-to-end enhancement networks is the autoencoder which maps input signals into an embedding space which is effective at separating speech and noise components. Some prior studies have relied on mappings such as the Short-Time Fourier Transform (STFT) or Discrete Cosine Transform (DCT) \cite{Williamson2015,Soni2018,Koizumi2018,Tan2019,Xu2020,Li2021b}, while others have explored trainable encoders and decoders \cite{Luo2019,Casebeer2020,Borgstrom2021}.
\par
The human cochlea is well known to implement a complex set of auditory filters that reflect a broad trade-off between frequency precision and temporal resolution \cite{olshausen2002}, extracting a multiscale representation of speech signals. The resulting filterbank conveys a non-uniform time-frequency tiling with a general emphasis on frequency resolution at the lower end of the auditory range and finer temporal resolution at the upper end. For decades speech scientists have successfully exploited the ear's frequency characteristics to great effect. The input features to the vast majority of speech-derived analysis methods include some form of frequency scale manipulation (e.g. mel and bark scale warping) that emulates the cochlea's spectral properties. Similarly, but to a lesser degree, the ear's temporal characteristics have been mimicked via a multiplicity of analysis window durations that allow for the detection of transient events without sacrificing the required frequency resolution. The narrow-band and wide-band spectrogram\cite{oppenheim1970speech} represent two extremes of time-frequency trade-off, while methods such as the multi-resolution short-time Fourier Transform\cite{smith2011}, Constant-Q transform\cite{brown1991}, and wavelet filtering \cite{mallat2008wavelet} implement filtering schemes that reflect the cochlea's temporal-spectral properties to some degree.
\par
In the field of deep learning, several recent studies have explored neural network topologies with multiscale representations \cite{Xie2017,huang2018,yu2019,Desplanques2020}. Such approaches typically aggregate representations across network layers in order to simultaneously model information at different scales. Multiscale network architectures have been shown highly successful for tasks such as image classification and segmentation \cite{Xie2017,huang2018,yu2019}. Additionally, \cite{Desplanques2020} and \cite{villalba22} showed the effectiveness of multiscale networks in modeling speech for the task of speaker verification.
\par
This paper proposes the multiscale autoencoder (MSAE) for mask-based end-to-end neural network speech enhancement, which is motivated both by recent multiscale network topologies and traditional multiresolution transforms in speech processing. The framework provides the encoder and decoder mappings required by mask-based approaches, and is not specific to the architecture used for mask estimation. The MSAE performs spectral decomposition of an input waveform within separate band-limited branches, each operating with a different rate and scale, to extract a sequence of embeddings. The proposed framework features intuitive parameterization of the autoencoder, including a flexible spectral band design based on the Constant-Q transform. Additionally, the MSAE is constructed entirely of differentiable operators, allowing it to be implemented within an end-to-end neural network, and discriminatively trained for the task of speech enhancement.
\par
To assess the performance of the proposed MSAE framework, it is combined with an example mask estimator based on the $U$-Net architecture \cite{Ronneberger2015} to form the MSAE-UNet end-to-end enhancement system. The paper provides experimental results comparing the performance of MSAE-UNet to several state-of-the-art methods in terms of objective speech quality metrics, and shows the proposed system to significantly outperform them. Additionally, the paper presents experimental results for automatic speech recognition using enhancement as a pre-processing step, and again clearly shows the benefit of the proposed framework.
\par 
This paper is organized as follows. Section \ref{sec:bnet} provides an overview of mask-based end-to-end speech enhancement systems. The MSAE framework is presented in Section \ref{sec:multiscale}. Section \ref{sec:mask_estimation} then introduces an example mask estimation architecture, that when integrated in the proposed autoencoder framework forms the end-to-end MSAE-UNet. Experimental results are presented in \ref{sec:results} and Section \ref{sec:conclusion} provides conclusions.
\section{Mask-Based End-to-end Enhancement}
\label{sec:bnet}
Let $\s\in\mathbb{R}^{\D}$ be a clean speech waveform, where
\begin{align}
\s=\left[s\left(1\right),\ldots,s\left(\D\right)\right]^{T}.
\end{align}
Let an observed speech waveform, $\x\in\mathbb{R}^{\D}$, be similarly defined. Here, the observed signal is captured in a real-world environment, and may suffer from distortions such as additive noise and reverberation. In this paper, speech enhancement is defined as the task of inferring the clean waveform, $\s$, from the observed version, $\x$, yielding the estimate $\sh\in\mathbb{R}^{\D}$.
\par
\begin{figure}
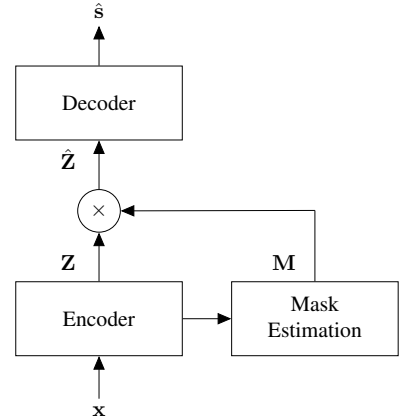

\centering
\scalebox{0.82}{
  \tikz{
\node at (0.0,0.0) [minimum width=0.75cm] (input) {$\x$};
\node at (0.0,1.5) [shape=rectangle,minimum height=1.2cm,minimum width=2.7cm,draw] (enc) {Encoder};
\draw [->] (input.north) -- (enc.south);
\node at (3.5,1.5) [shape=rectangle,minimum height=1.2cm,minimum width=2.7cm,draw] (mask) {\begin{tabular}{c} Mask \\ Estimation \end{tabular}};
\draw [->] (enc.east) -- (mask.west);
\node at (0.0,3.25) [shape=circle,minimum size=0.5cm,draw] (mult) {$\times$};
\draw [->] (enc.north) -- (mult.south);
\draw [-] (mask.north) -- ($ (mask.north) + (0.0,1.15) $);
\draw [->] ($ (mask.north) + (0.0,1.15) $) -- (mult.east);
\node at (0.0,5.0) [shape=rectangle,minimum height=1.2cm,minimum width=2.7cm,draw] (dec) {Decoder};
\draw [->] (mult.north) -- (dec.south);
\node at (0.0,6.5) [minimum width=0.75cm] (output) {$\sh$};
\draw [->] (dec.north) -- (output.south);
\node at ($ (enc.north) + (-0.5,0.3) $) [minimum width=0.75cm] (z) {$\z$};
\node at ($ (mask.north) + (-0.5,0.3) $) [minimum width=0.75cm] (z) {$\m$};
\node at ($ (dec.south) + (-0.5,-0.3) $) [minimum width=0.75cm] (z) {$\zh$};
 }
 }
 \caption{Overview of the \textit{b}-Net Architecture \cite{Borgstrom2021}. Note that with the masking mechanism disabled,  the \textit{b}-Net results in an autoencoder.}
 \label{fig:bnet}
\end{figure}

 
This paper focuses on a mask-based approach to end-to-end neural network speech enhancement. Specifically, we use the \textit{b}-Net framework from \cite{Borgstrom2021}, illustrated in Figure \ref{fig:bnet}, and named for its likeness to the lower-case letter. The general network architecture is delineated by its application of a multiplicative mask in a general, possibly learned, embedding space.  Similar networks have been explored in the context of source separation \cite{Venkataramani2018,Luo2019} as well as signal enhancement \cite{Borgstrom2021}. The \textit{b}-Net structure can be interpreted as a generalization of statistical model-based  speech  enhancement  methods \cite{Mcaulay1980,Ephraim1984,Ephraim1985,Cohen2002}. With these earlier systems the short-time magnitude spectrogram is extracted from the noisy input waveform, manipulated via a multiplicative mask, and the output waveform is then generated from the enhanced spectrogram through overlap-and-add synthesis using the original noisy phase signal. With \textit{b}-Net, the Fourier analysis and overlap-and-add synthesis are replaced by generalized encoder and decoder mappings.
\par 
Within the \textit{b}-Net framework, as can be observed in Figure \ref{fig:bnet}, an encoder network extracts a set of embeddings from the input according to the generalized mapping
\begin{align}
\label{eqn:encoder}
\z=\fenc\left(\x\right),
\end{align}
where $\z$ is a tensor of size $T\times K\times C$. Here, the first axis corresponds to time, and $T$ denotes the number of frames extracted from the input waveform. The second axis corresponds to frequency, and $K$ denotes the number of bins used in the frequency representation. Finally, $C$ represents the number of channels in the feature map $\z$. Enhancement is performed via multiplicative masking in the embedding space defined by $\fenc$, so that interfering signal components are appropriately attenuated. A mask estimation network extracts this multiplicative mask according to   
\begin{align}
\label{eqn:mask}
\m=\fmask\left(\z\right),
\end{align}
where $\m$ is a tensor of the same size as $\z$, and contains values in the range $\left[0,1\right]$. 
Enhanced versions of the embeddings are obtained via element-wise multiplication according to
\begin{align}
\zh=\m\otimes\z,
\end{align}
where $\otimes$ denotes the Kronecker product. Finally, a decoder network extracts the output waveform from the enhanced embeddings according to the generalized mapping
\begin{align}
\sh&=\fdec\left(\zh\right).
\end{align}
Note that the overall operation of the \textit{b}-Net framework can be expressed concisely as
\begin{align}
\label{eqn:bnet}
\sh=\fdec\left(\fmask\left(\fenc\left(\x\right)\right)\otimes\fenc\left(\x\right)\right).
\end{align}
\par 
A key feature of the \textit{b}-Net architecture is the ability to disable the masking mechanism. This results in an autoencoder path expressed as
\begin{align}
\label{eqn:autoencoder}
\xh=\fdec\left(\fenc\left(\x\right)\right),
\end{align}
where $\xh$ is the reconstructed version of the input $\x$. Typically the autoencoder path is able to produce an accurate reconstruction of the input, leading to the approximation 
\begin{align}
\label{eqn:autoencoder_approx}
\x\approx\fdec\left(\fenc\left(\x\right)\right).
\end{align}
Note that in general, end-to-end architectures do not contain an analogous autoencoder path, such as in Fully Convolutional Networks (FCNs) \cite{Fu2018,Pandey2019,Giri2019}. The existence of this path allows the user to dynamically control the level of noise suppression via a minimum gain level by adapting (\ref{eqn:bnet}) so that
\begin{align}
\sh=\fdec\left(\max\left\{\gmin,\fmask\left(\fenc\left(\x\right)\right)\right\}\otimes\fenc\left(\x\right)\right),
\end{align}
where $\gmin\in\left[0,1\right]$ can be tuned during inference \cite{Borgstrom2021}. In this way, the user can efficiently navigate the inherent trade-off between noise suppression and speech quality. In \cite{Borgstrom2021}, it was shown that dynamic control of the level of noise suppression can provide benefits in enhancement performance, both for objective speech quality metrics and for subjective listening tests.
\par 
The focus of this paper is the design of the autoencoder path defined in (\ref{eqn:autoencoder}). Several studies have used the Short-Time Fourier Transform (STFT) to define encoder and decoder mappings \cite{Williamson2015,Soni2018,Tan2019,Xu2020}, while other work has relied on the Discrete Cosine Transform (DCT) \cite{Koizumi2018,Li2021b}. Finally, trainable $1$-dimensional Convolutional Neural Network (CNN) layers have showed promise for learning an embedding space for effective masking-based speech enhancement \cite{Luo2019,Casebeer2020,Borgstrom2021}. In the following section, the MSAE is developed to provide a multirate and multiscale embedding space for mask-based speech enhancement.
\section{The Multiscale Autoencoder}
\label{sec:multiscale}
This section presents the Multiscale Autoencoder. First, definitions are provided for the analysis operator, $\Aop$, and synthesis operator, $\Sop$, which represent the basic building blocks of the MSAE. Next, the encoder network is described, which performs spectral decomposition of an input waveform within separate band-limited branches, each operating with a different rate and scale. Finally, the decoder network is described, which synthesizes the output waveform from the encoder output. Note that the encoder and decoder networks can be constructed entirely using standard neural network operations, allowing discriminative training of autoencoder parameters. 
\subsection{The Analysis Operator}
\label{sec:analysis}
The analysis operator, $\Aop$, performs a band-limited spectral decomposition of the input waveform $\x$, outputting a tensor $\y$. It is parameterized by the tuple $\left(N,\wo,\wt\right)$, were $N$ is the duration of the analysis window in samples, and $\left[\wo,\wt\right]$ is the range of normalized frequencies included in the analysis.
\par
The analysis operator relies on an invertible transform, either fixed or learned, to perform spectral analysis. The DFT represents one possible instantiation of the analysis operator and is used in this study, but other transforms, e.g. the DCT, can also be used. Let $\F=\left[\fo,\ldots,\fN\right]$ be the $N$-dimensional DFT transform. Define $\W$ to contain the basis vectors of $\F$ which correspond to the intended spectral range $\left[\wo,\wt\right]$, i.e.
\begin{align}
\label{eqn:W}
\W=\left[\mathbf{f}_{\lceil N\wo/2\rceil},\ldots,\mathbf{f}_{\lfloor N\wt/2\rfloor}\right]\in\mathbb{R}^{N\times K_{W}},
\end{align}
where $K_{W}=\lfloor N\wt/2\rfloor-\lceil N\wo/2\rceil+1$. To perform spectral analysis, the $\Aop$ operator applies $4$ parallel $1$-dimensional CNN layers, using the kernels $\mathfrak{Re}\left(\W\right)$, $\mathfrak{Re}\left(-\W\right)$, $\mathfrak{Im}\left(\W\right)$, and $\mathfrak{Im}\left(-\W\right)$, respectively. Each layer uses a stride of $N/2$ and Rectified Linear Unit (ReLU) activation functions, resulting in parallel tensors of shape $\lfloor 2\D/N\rfloor\times K_{W}$. 
The outputs of the CNN layers are then concatenated to form the tensor $\y$ with shape $\lfloor 2\D/N\rfloor\times K_{W}\times4$. Since $\x$ is a real-valued signal, its spectrum is conjugate-symmetric. For the sake of computational efficiency, the kernel $\W$ in (\ref{eqn:W}) therefore only contains DFT basis functions in the normalized frequency range $\left[0,1\right]$, since the remainder of the spectrum can be reconstructed from these basis functions.
\par 
Note that the motivation for using $4$ CNN layers is to enable representation of complex numbers as a vector of non-negative values. However, for the sake of efficiency, the $4$ CNN layers can be implemented as $2$ CNN layers with kernels $\mathfrak{Re}\left(\W\right)$ and $\mathfrak{Im}\left(\W\right)$. The outputs can then be expanded to $4$ parallel layers by separately considering non-negative and negative values of each, prior to the ReLU activations.
\par 
For improved spectral properties, windowing functions can be leveraged when constructing the kernel in (\ref{eqn:W}). That is, a window $\h\in\mathbb{R}^{N}$ can be applied to the DFT basis functions as $\fk\leftarrow\h\otimes\fk$. Note that the chosen windowing function should provide unity overlap-and-add when used in conjunction with the synthesis operator \cite{Quatieri2008}. In this study, the square-root of the Hanning window is used.
\subsection{The Synthesis Operator}
\label{sec:synthesis}
The synthesis operator, $\Sop$, generates a band-limited waveform $\x$ from the tensor $\y$. It applies a $1$-dimensional Transpose CNN layer to each of the $4$ slices in $\y$, using the corresponding kernels defined in Section \ref{sec:analysis}. A stride of $N/2$ is used for each layer, and the outputs are summed to form $\x$. 
Note that when identically parameterized, the analysis operator inverts the synthesis operator, i.e.
\begin{align}
\label{eqn:invertible}
\y=\Aop\left(\Sop\left(\y;N,\wo,\wt\right);N,\wo,\wt\right).
\end{align}
Note that $\W$ in (\ref{eqn:W}) only contains DFT basis functions within the normalized frequency range $\left[0,1\right]$, omitting the other half of the spectrum. Since $\x$ is a real-valued signal, its spectrum is conjugate-symmetric, and the spectrum can be reconstructed by simply scaling certain DFT basis functions during synthesis. Specifically, when constructing $\W$ in the synthesis operator, all basis functions $\fk$ are scaled by $2$, except for those corresponding to the DC and Nyquist frequencies. 
\subsection{The Encoder Network}
\label{sec:encoder}
Having defined the analysis and synthesis operators, the encoder and decoder networks can be presented. The encoder performs spectral decomposition of an input waveform $\x$ within separate band-limited branches, each operating with a different rate and scale, and outputs a tensor, $\z$, providing the mapping in (\ref{eqn:encoder}). The network relies on the analysis operator $\Aop$ for spectral processing. The encoder network is parameterized by the number of branches, $B$, and the duration of the base analysis window in samples, $\No$. Additionally, the encoder network is parameterized by the normalized frequency ranges which split the input spectrum into $B$ roughly orthogonal bands, requiring the $B+1$ band edges $\wb$ for $0\leq b\leq B$.
\par
\begin{figure*}
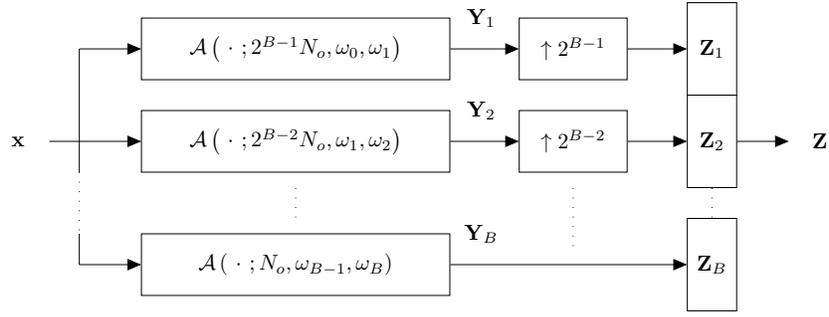

\centering
\scalebox{0.82}{
  \tikz{
\node at (-4.5,0.5) [minimum width=1cm] (input) {$\x$};
\node at (0,2) [shape=rectangle,minimum height=1cm,minimum width=5cm,draw] (cnn1) {$\Aop\left(\ \cdot\ ;2^{B-1}\No,\wz,\wo\right)$};
\node at (0,0.5) [shape=rectangle,minimum height=1cm,minimum width=5cm,draw] (cnn2) {$\Aop\left(\ \cdot\ ;2^{B-2}\No,\wo,\wt\right)$};
\node at (0,-1.5) [shape=rectangle,minimum height=1cm,minimum width=5cm,draw] (cnnB) {$\Aop\left(\ \cdot\ ;\No,\wBmo,\wB\right)$};

\node at (4.5,2) [shape=rectangle,minimum height=1cm,minimum width=1.75cm,draw] (up1) {$\uparrow 2^{B-1}$};
\node at (4.5,0.5) [shape=rectangle,minimum height=1cm,minimum width=1.75cm,draw] (up2) {$\uparrow 2^{B-2}$};

\node at (6.75,2) [shape=rectangle,minimum height=1.5cm,minimum width=0.8cm,draw] (z1) {$\zo$};
\node at (6.75,0.5) [shape=rectangle,minimum height=1.5cm,minimum width=0.8cm,draw] (z2) {$\zt$};
\node at (6.75,-1.5) [shape=rectangle,minimum height=1.5cm,minimum width=0.8cm,draw] (zB) {$\zB$};

\node at ($ (cnn1.east) + (0.5,0.5) $) [] (y1) {$\yo$};
\node at ($ (cnn2.east) + (0.5,0.5) $) [] (y2) {$\yt$};
\node at ($ (cnnB.east) + (0.5,0.5) $) [] (yB) {$\yB$};

\draw [->] ($ (cnn1.west) + (-1,0) $) -- (cnn1.west);
\draw [->] (input.east) -- (cnn2.west);
\draw [->] ($ (cnnB.west) + (-1,0) $) -- (cnnB.west);

\draw [-] ($ (cnn1.west) + (-1,0) $) -- ($ (cnn2.west) + (-1,-0.5) $);
\draw [-] ($ (cnnB.west) + (-1,0.5) $) -- ($ (cnnB.west) + (-1,0) $);
\draw[loosely dotted] ($ (cnn2.west) + (-1,-0.5) $) -- ($ (cnnB.west) + (-1,0.5) $);

\draw [->] (cnn1.east) -- (up1.west);
\draw [->] (cnn2.east) -- (up2.west);

\draw [->] (up1.east) -- (z1.west);
\draw [->] (up2.east) -- (z2.west);
\draw [->] (cnnB.east) -- (zB.west);

\draw[loosely dotted] ($ (cnn2.south) + (0,-0.25) $) -- ($ (cnnB.north) +  (0,0.25) $);
\draw[loosely dotted] ($ (up2.south) + (0,-0.25)$) -- ($ (up2.south) + (0,-1.25) $);
\draw[loosely dotted] (z2.south) -- (zB.north);

\node at (8.5,.5) [minimum width=1cm] (output) {$\z$};
\draw [->] (z2.east) -- (output.west);
 }
 }
 \caption{The Encoder Network of the MSAE: The analysis operator $\Aop$ is defined in Section \ref{sec:analysis}, the $\uparrow K$ operator denotes nearest-neighbor upsampling in time by rate $K$, and stacked blocks denote tensor concatenation along the $2^{nd}$ axis.}
 \label{fig:encoder}
\end{figure*}

 
Figure \ref{fig:encoder} provides an overview of the encoder network. The input waveform $\x$ is first processed by $B$ parallel analysis operators, resulting in spectral decomposition tensors $\yb$, each operating at a different frame rate, and each of shape
\begin{align}
&\lfloor 2^{-\left(B-b-1\right)}\D/\No\rfloor\\\nonumber
&\ \ \ \times\left(\lfloor 2^{B-b-1}\No\wb\rfloor-\lceil 2^{B-b-1}\No\wbmo\rceil+1\right)\times 4.   
\end{align}
Next, tensors $\yb$ are upsampled by rate $2^{B-b}$ in time, via frame repetition, resulting in tensors $\zb$ all operating at the same base frame rate. Finally, tensors $\zb$ are concatenated along the $2^{nd}$ axis, resulting in the output tensor $\z$, containing a multi-scale spectral decomposition of input $\x$. The shape of $\z$ is $\lfloor 2\D/\No\rfloor\times K_{T}\times4$, where
\begin{align}
\label{eqn:channels}
K_{T}=B+\sum^{B}_{b=1}\left(\lfloor 2^{B-b-1}\No\wb\rfloor-\lceil 2^{B-b-1}\No\wbmo\rceil\right).
\end{align}
\subsection{The Decoder Network}
\label{sec:decoder}
\begin{figure*}
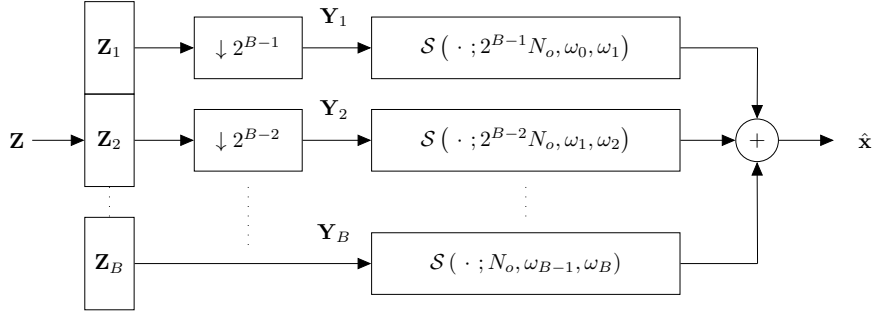

\centering
\scalebox{0.82}{
  \tikz{
\node at (-4.5,0.5) [] (z) {$\z$};
\node at (-3,2) [shape=rectangle,minimum height=1.5cm,minimum width=0.8cm,draw] (z1) {$\zo$};
\node at (-3,0.5) [shape=rectangle,minimum height=1.5cm,minimum width=0.8cm,draw] (z2) {$\zt$};
\node at (-3,-1.5) [shape=rectangle,minimum height=1.5cm,minimum width=0.8cm,draw] (zB) {$\zB$};
\draw[loosely dotted] (z2.south) -- (zB.north);
\draw [->] (z.east) -- (z2.west);
\node at (-0.75,2) [shape=rectangle,minimum height=1cm,minimum width=1.75cm,draw] (dn1) {$\downarrow 2^{B-1}$};
\node at (-0.75,0.5) [shape=rectangle,minimum height=1cm,minimum width=1.75cm,draw] (dn2) {$\downarrow 2^{B-2}$};
\draw [->] (z1.east) -- (dn1.west);
\draw [->] (z2.east) -- (dn2.west);
\node at (3.75,2) [shape=rectangle,minimum height=1cm,minimum width=5cm,draw] (cnn1) {$\Sop\left(\ \cdot\ ;2^{B-1}\No,\wz,\wo\right)$};
\node at (3.75,0.5) [shape=rectangle,minimum height=1cm,minimum width=5cm,draw] (cnn2) {$\Sop\left(\ \cdot\ ;2^{B-2}\No,\wo,\wt\right)$};
\node at (3.75,-1.5) [shape=rectangle,minimum height=1cm,minimum width=5cm,draw] (cnnB) {$\Sop\left(\ \cdot\ ;\No,\wBmo,\wB\right)$};
\node at ($ (cnn1.west) + (-0.6,0.5) $) [] (y1) {$\yo$};
\node at ($ (cnn2.west) + (-0.6,0.5) $) [] (y2) {$\yt$};
\node at ($ (cnnB.west) + (-0.6,0.5) $) [] (yB) {$\yB$};
\draw [->] (dn1.east) -- (cnn1.west);
\draw [->] (dn2.east) -- (cnn2.west);
\draw [->] (zB.east) -- (cnnB.west);
\draw[loosely dotted] ($ (cnn2.south) + (0,-0.25) $) -- ($ (cnnB.north) +  (0,0.25) $);
\draw[loosely dotted] ($ (dn2.south) + (0,-0.25)$) -- ($ (dn2.south) + (0,-1.25) $);
\node at (7.5,0.5) [shape=circle,minimum size=0.5cm,draw] (sum) {$+$};
\node at (9.25,0.5) [minimum size=1cm] (output) {$\xh$};
\draw [->] (cnn2.east) -- (sum.west);
\draw [->] (sum.east) -- (output.west);
\draw [-] (cnn1.east) -- ($ (cnn1.east) + (1.25,0) $);
\draw [->] ($ (cnn1.east) + (1.25,0) $) -- (sum.north);
\draw [-] (cnnB.east) -- ($ (cnnB.east) + (1.25,0) $);
\draw [->] ($ (cnnB.east) + (1.25,0) $) -- (sum.south);
 }
 }
 \caption{The Decoder Network of the MSAE: The synthesis operator $\Sop$ is defined in Section \ref{sec:synthesis}, the $\downarrow K$ operator denotes Max-Pooling in time with pool size $K$ and stride $K$, and stacked blocks denote tensor concatenation along the $2^{nd}$ axis.}
 \label{fig:decoder}
\end{figure*}

The decoder network, illustrated in Figure \ref{fig:decoder}, synthesizes the output waveform $\xh$ from the encoder output $\z$, according to (\ref{eqn:bnet}). First, $\z$ is split into branch-specific tensors $\zb$. Next, Max-Pooling in time is applied to each, with pool size $2^{B-b}$ and stride $2^{B-b}$, yielding tensors $\yb$. Finally, the synthesis operator $\Sop$ is utilized to generate the reconstructed waveform according to 
\begin{align}
\label{eqn:decoder}
\xh=\sum^{B}_{b=1}\Sop\left(\yb;2^{B-b}\No,\wbmo,\wb\right).
\end{align}
Note that besides the trivial case of $B=1$, the MSAE does not provide perfect reconstruction, which is due to adjacent DFT basis functions at branch edges having different dimensions and therefore not being strictly orthogonal. However, distortion due to reconstruction error was deemed perceptually negligible for MSAE configurations used during experimentation, leading to the approximation in (\ref{eqn:autoencoder_approx}).
\subsection{The Design of Spectral Bands}
\label{sec:quality_factor}
The spectral bands of the MSAE can be designed in various ways. For example, the spectrum can be split into $B$ bands of uniform width. In this paper, however, we use a generalized Constant-Q design \cite{brown1991}, where the quality factor $\Q$ is defined as the ratio of band center frequency to band width,
\begin{align}
\label{eqn:quality_factor1}
Q=\frac{\wb+\wbmo}{2\left(\wb-\wbmo\right)},
\end{align}
and is constant across bands. Rearranging (\ref{eqn:quality_factor1}) leads to the recursive expression
\begin{align}
\label{eqn:quality_factor2}
\wbmo=\rho^{-1}\wb,
\end{align}
where
\begin{align}
\rho=\frac{2\Q+1}{2\Q-1},
\end{align}
for $\Q>0.5$. 
If the uppermost band edge is set to the Nyquist frequency, i.e. $\wB=1$, then (\ref{eqn:quality_factor2}) leads to the band edge design
\begin{align}
\label{eqn:quality_factor3}
\wb=\left\{\begin{array}{ll}
	    \rho^{-B+b} & \textrm{if}\ \  0<b\leq B\\
	    0 & \textrm{if}\ \  b=0\end{array}\right..
\end{align}
Note that if $\Q=1.5$, then $\rho=2$, representing a dyadic decomposition.
\par 
Using the Constant-Q band design, MSAE configurations can be fully parameterized by the tuple ($B$,$\Q$,$\No$). However, in the remainder of the paper signals will be assumed to have a sampling frequency of $\Fs=16$kHz, and configurations will instead be specified by the tuple ($B$,$\Q$,$\To$), where $\To=\No/\Fs$, which allows for more intuition when comparing MSAE configurations. Additionally, for MSAE systems with $B=1$, the band design is irrelevant, and the $-$ symbol will be used to specify quality factor.
\par
\begin{figure*}
	\centering
	\includegraphics[trim={3.5cm 0cm 2.2cm 0cm},width=7.0in]{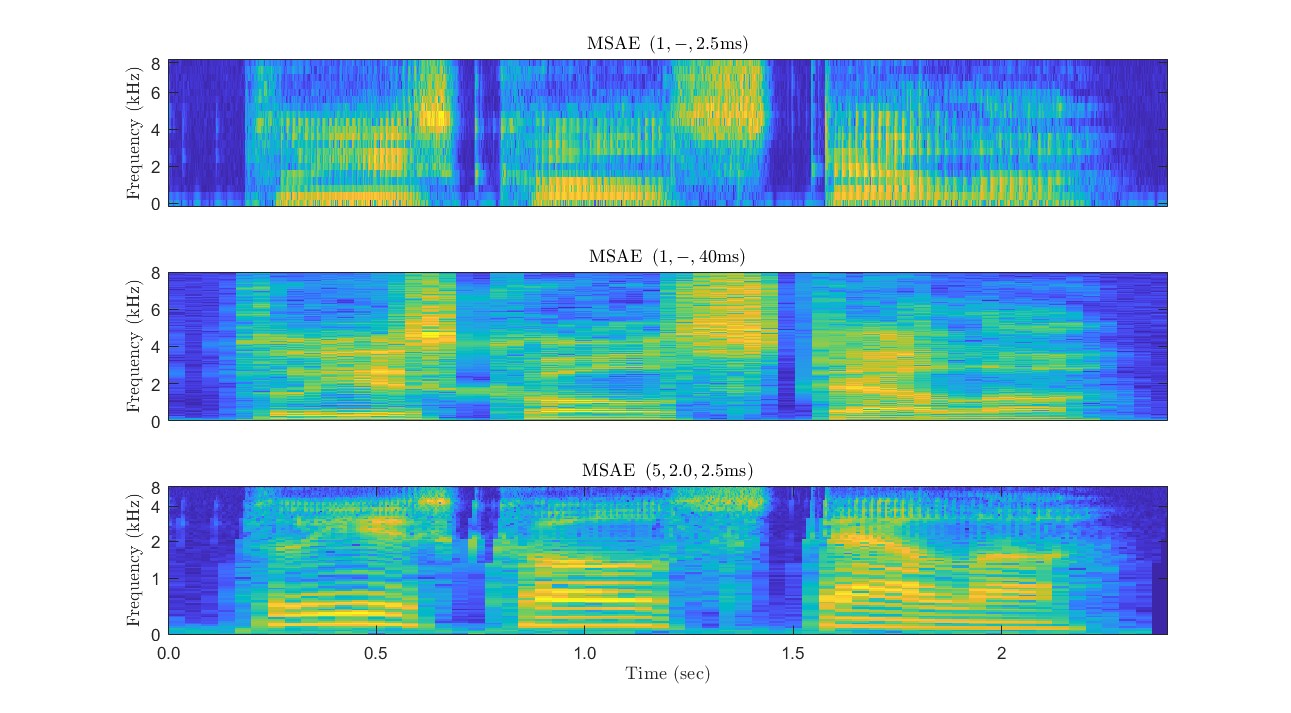}
	\caption{Magnitude spectral representations of $\z$ for various MSAE configurations, for a clean speech signal from the VCTK Noisy and Reverberant set \cite{vctk} with the transcription "Please call Stella": The top panel shows the $\left(1,-,2.5\textrm{ms}\right)$ configuration, which corresponds to conventional wideband processing with an analysis window duration of $2.5$ms. The second panel shows the $\left(1,-,40\textrm{ms}\right)$ configuration, which corresponds to conventional narrowband processing with an analysis window duration of $40$ms. Finally, the bottom panel provides the $\left(5,2.0,2.5\textrm{ms}\right)$ configuration, which operates with five bands spanning analysis window durations between $2.5$ms and $40$ms.}
	\label{fig:spectrograms}
\end{figure*}

The embedding space defined by the MSAE encoder differs from conventional time-frequency representations with respect to two main characteristics. First, the multiscale processing of the MSAE provides the ability to capture improved spectral resolution in lower frequency bands, while simultaneously achieving greater temporal resolution in higher frequency regions. Second, since the longer analysis windows used to model low frequency regions provide higher spectral resolution, the time-frequency representation defined by the MSAE naturally includes a warped frequency axis with more emphasis applied to lower frequency bands. A magnitude spectral representation of the tensor $\z$ can be derived by calculating the root-mean-square (RMS) along the $3^{rd}$ axis, yielding a $T\times K$ matrix. Figure \ref{fig:spectrograms} provides example magnitude spectral representations of $\z$ for various MSAE configurations, for a clean speech signal from the VCTK Noisy and Reverberant set \cite{vctk} with the transcription "Please call Stella." The top panel shows the $\left(1,-,2.5\textrm{ms}\right)$ configuration, which corresponds to conventional wideband processing with an analysis window duration of $2.5$ms. The second panel shows the $\left(1,-,40\textrm{ms}\right)$ configuration, which corresponds to conventional narrowband processing with an analysis window duration of $40$ms. Finally, the bottom panel provides the $\left(5,2.0,2.5\textrm{ms}\right)$ configuration, which corresponds to five bands spanning analysis window durations between $2.5$ms and $40$ms.
\par
\begin{figure}
	\centering
	\includegraphics[trim={0.5cm 0cm 0.5cm 0cm},width=3.4in]{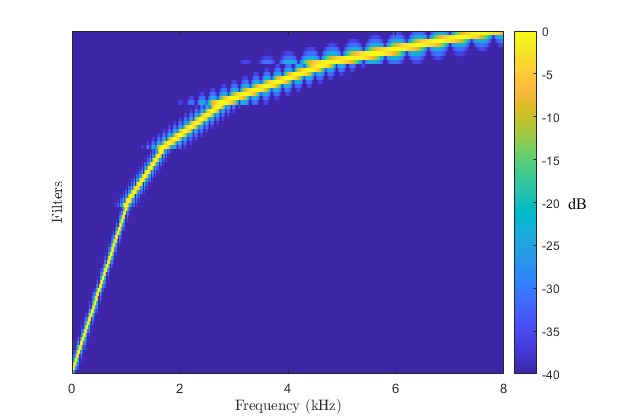}
	\caption{Frequency responses of the $1$-dimensional CNN filters utilized by the MSAE with $\left(5,2.0,2.5\textrm{ms}\right)$ configuration.}
	\label{fig:freq_responses}
\end{figure}

 
As can be observed in Figure \ref{fig:spectrograms}, the wideband time-frequency representation of the top panel captures fine temporal patterns such as the plosives $/k/$ and $/t/$ at $0.75$s and $1.5$s, respectively. Conversely, the narrowband representation of the second panel achieves better resolution of harmonics and formant frequencies, e.g. the $/a/$ vowel at $2.0$s. The multiscale representation in the bottom panel, however, achieves both wideband representation in higher frequency regions, and narrowband representation in lower frequency regions. Additionally, it can be observed that the multiscale embedding space in the bottom panel results in a warped frequency scale. This frequency warping can also be observed in Figure \ref{fig:freq_responses}, which illustrates the frequency responses of the $1$-dimensional CNN filters utilized by the $\left(5,2.0,2.5\textrm{ms}\right)$ MSAE. In Figure \ref{fig:freq_responses}, it can be observed that a greater number of CNN filters is allocated to lower frequency ranges. Additionally, due to the longer analysis windows applied in the lower frequency bands, the corresponding filters show a narrower passband.
\subsection{The MSAE as a Trainable Network}
\label{sec:learned_filters}
The MSAE encoder and decoder networks are composed of differentiable operations, and can be constructed entirely using standard neural network blocks. This allows the kernels $\W$ to be trained jointly with the mask estimation network for the task of speech enhancement. Note that the kernels can be initialized according to (\ref{eqn:W}), and allowed to adapt during the overall network training process. Additionally, the tensors $\W$ can be overcomplete, i.e. $K_{W}>\lfloor N\wt/2\rfloor-\lceil N\wo/2\rceil+1$, providing greater modeling capacity in the spectral decomposition performed by the encoder network. In this case, an overcompleteness factor, $\overcomp$, is defined such that
\begin{align}
    K_{W}=\lfloor\overcomp\left(\lfloor N\wt/2\rfloor-\lceil N\wo/2\rceil+1\right)\rfloor.
\end{align}
In this study, for $\overcomp>1$, the tensor $\W$  was initialized with DFT basis vectors with interpolated center frequencies.
\section{Mask Estimation}
\label{sec:mask_estimation}
The role of the mask estimator within the $b$-Net framework is to map $\z$ to a tensor $\m$ via (\ref{eqn:mask}), which is used as a multiplicative mask to perform speech enhancement in the embedding space defined by the encoder. The proposed MSAE framework is not specific to the architecture of the mask estimation network, and is instead compatible with the generalized mapping given in (\ref{eqn:mask}). However, in order to assess the performance of the MSAE framework, this study leverages an example mask estimation network based on the $U$-Net architecture from \cite{Ronneberger2015}, which is described in this section. During experimentation, when the example \textit{U}-Net mask estimation architecture is combined with the proposed MSAE framework, the resulting end-to-end enhancement system is referred to as the MSAE-UNet.
\par
\begin{figure}
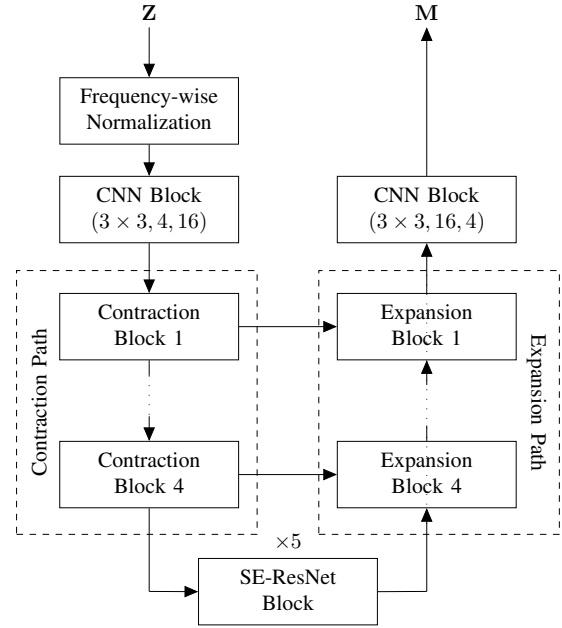

\centering
\scalebox{0.82}{
  \tikz{
\node at (0.0,-1.3) [minimum width=0.75cm] (input) {$\z$};
\node at (0.0,-2.9) [shape=rectangle,minimum height=1.0cm,minimum width=2.8cm,draw] (cep) {\begin{tabular}{c}Frequency-wise\\Normalization\end{tabular}};
\node at (0.0,-4.5) [shape=rectangle,minimum height=1.0cm,minimum width=2.9cm,draw] (cnn3) {\begin{tabular}{c}CNN Block\\$\left(3\times3,4,16\right)$ \end{tabular}};
\node at (0.0,-6.4) [shape=rectangle,minimum height=1.0cm,minimum width=2.9cm,draw] (ds1) {\begin{tabular}{c} Contraction\\ Block 1\end{tabular}};
\node at (0.0,-8.8) [shape=rectangle,minimum height=1.0cm,minimum width=2.9cm,draw] (ds2) {\begin{tabular}{c} Contraction\\ Block 4\end{tabular}};
\node at (4.5,-4.5) [shape=rectangle,minimum height=1.0cm,minimum width=2.9cm,draw] (cnn2) {\begin{tabular}{c} CNN Block \\ $\left(3\times3,16,4\right)$ \end{tabular}};
\node at (4.5,-6.4) [shape=rectangle,minimum height=1.0cm,minimum width=2.9cm,draw] (us2) {\begin{tabular}{c} Expansion\\ Block 1\end{tabular}};
\node at (4.5,-8.8) [shape=rectangle,minimum height=1.0cm,minimum width=2.9cm,draw] (us1) {\begin{tabular}{c} Expansion\\ Block 4\end{tabular}};
\node at (4.5,-1.3) [minimum width=0.75cm] (output) {$\m$};
\node at (2.25,-10.7) [shape=rectangle,minimum height=1.0cm,minimum width=2.9cm,draw] (base) {\begin{tabular}{c} SE-ResNet\\ Block \end{tabular}};
\node at (2.25,-9.9) [minimum width=0.75cm] (times) {$\times5$};
\node at (-0.2,-7.63) [minimum width=3.9cm,minimum height=4.28cm,draw,dashed] (exp) {};
\node at (4.7,-7.63) [minimum width=3.9cm,minimum height=4.28cm,draw,dashed] (cont) {};
\node at (-1.78,-7.63) [rotate=90] (contlab) {Contraction Path};
\node at (6.28,-7.63) [rotate=-90] (explab) {Expansion Path};
\draw [->] (input.south) -- (cep.north);
\draw [->] (cep.south) -- (cnn3.north);
\draw [->] (cnn3.south) -- (ds1.north);
\draw [->] (cnn2.north) -- (output.south);
\draw [->] (us2.north) -- (cnn2.south);
\draw [->] (ds1.east) -- (us2.west);
\draw [->] (ds2.east) -- (us1.west);
\draw [-] (ds2.south) -- ($ (ds2.south) + (0.0,-1.35) $);
\draw [->] ($ (ds2.south) + (0.0,-1.35) $) -- (base.west);
\draw [->] ($ (us1.south) + (0.0,-1.35) $) -- (us1.south);
\draw [-] (base.east) -- ($ (us1.south) + (0.0,-1.35) $);
\draw [-] (ds1.south) -- ($ (ds1.south) + (0.0,-0.25) $);
\draw [->] ($ (ds2.north) + (0.0,0.4) $) -- (ds2.north);
\draw [loosely dotted] ($ (ds1.south) + (0.0,-0.25) $) -- ($ (ds2.north) + (0.0,0.4) $);
\draw [-] (us1.north) -- ($ (us1.north) + (0.0,0.25) $);
\draw [->] ($ (us2.south) + (0.0,-0.4) $) -- (us2.south);
\draw [loosely dotted] ($ (us1.south) + (0.0,-0.25) $) -- ($ (us2.north) + (0.0,0.4) $);
 }
 }
 \caption{Overview of the Mask Estimation Network: $\z$ is the set of embeddings extracted from the input waveform via (\ref{eqn:encoder}), and $\m$ is the multiplicative mask used for enhancement in (\ref{eqn:bnet}).}
 \label{fig:mask}
\end{figure}

An overview of the mask estimation network used in this study is provided in Figure \ref{fig:mask}. As a basic component, the \textit{CNN Block} is defined as the following series of operations: a $2$-dimensional CNN layer, Batch Normalization \cite{Ioffe2015}, and non-linear activation functions which are ReLUs unless otherwise stated. This block is parameterized by the dimensions of the CNN filters, the number of input channels, and the number of output channels. Configurations of \textit{CNN Blocks} are expressed as tuples of these parameters, e.g. $\left(3\times3,32,64\right)$.
\par
As can be observed in Figure \ref{fig:mask}, the mask estimation network first applies Frequency-wise Normalization. This includes the element-wise $\log\left(\z+1\right)$ operator, followed by mean- and variance-normalization across time frames and channels, resulting in zero-mean and unit-variance feature map distributions per frequency bin. Frequency-wise Normalization is designed to promote invariance to domain shifts in the input signals, and is based on the Cepstral Extraction operator in \cite{Borgstrom2021}. A \textit{CNN Block} is then applied to expand the feature map to $16$ channels prior to processing by the \textit{U}-Net architecture,
\par
\begin{figure}
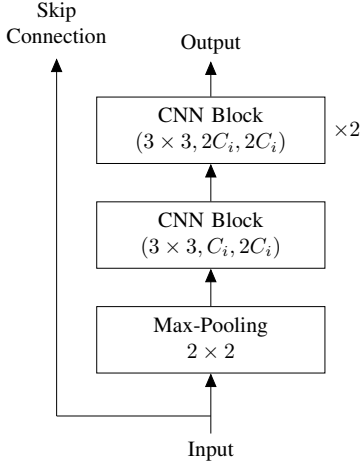

\centering
\scalebox{0.82}{
  \tikz{
\node at (0.0,-0.3) [minimum width=0.75cm] (input) {Input};
\node at (0.0,1.5) [shape=rectangle,minimum height=1.0cm,minimum width=3.7cm,draw] (ds) {\begin{tabular}{c} Max-Pooling \\ $2\times2$ \end{tabular}};
\node at (0.0,3.2) [shape=rectangle,minimum height=1.0cm,minimum width=3.7cm,draw] (cnn1) {\begin{tabular}{c} CNN Block\\$\left(3\times3,C_{i},2C_{i}\right)$ \end{tabular}};
\node at (0.0,4.9) [shape=rectangle,minimum height=1.0cm,minimum width=3.7cm,draw] (cnn2) {\begin{tabular}{c} CNN Block\\$\left(3\times3,2C_{i},2C_{i}\right)$ \end{tabular}};
\node at (2.2,4.9) [minimum width=0.75cm] (times) {$\times2$};
\draw [->] (input.north) -- (ds.south);
\draw [->] (ds.north) -- (cnn1.south);
\node at (0.0,6.3) [minimum width=0.75cm] (output) {Output};
\node at (-2.5,6.61) [minimum width=0.75cm] (output2) {\begin{tabular}{c} Skip\\ Connection\end{tabular}};
\draw [->] (cnn1.north) -- (cnn2.south);
\draw [->] (cnn2.north) -- (output.south);
\draw [->] ($ (output2.south) + (0.0,-5.8) $) -- (output2.south);
\draw [-] ($ (output2.south) + (0.0,-5.8) $) -- ($ (output2.south) + (2.5,-5.8) $);
 }
 }
 \caption{Overview of the Contraction Block: For a general Input shape of $T_{i}\times K_{i}\times C_{i}$, the Skip Connection has shape $T_{i}\times K_{i}\times C_{i}$ and the Output has shape $T_{i}/2\times K_{i}/2\times 2C_{i}$.}
 \label{fig:contraction}
\end{figure}

The mask estimation network next utilizes a \textit{U}-Net structure. As can be observed in Figure \ref{fig:mask}, a contraction path first applies a series of \textit{Contraction Blocks}, defined in Figure \ref{fig:contraction}, in order to increase the spectral and temporal context captured in the feature map. Max-Pooling using a $2\times2$ window size and a $2\times2$ stride is first applied, halving the dimensions of the feature map. A \textit{CNN Block} is then applied to double the number of channels, followed by two more \textit{CNN Blocks}. The \textit{Contraction Blocks} also provide a second output, which is a skip connection from the input. The contracting path includes $4$ levels, resulting in an output feature map with $256$ channels.
\par
\begin{figure}
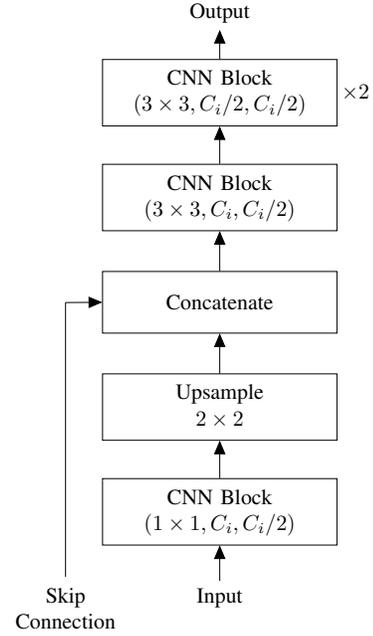

\centering
\scalebox{0.82}{
  \tikz{
\node at (0.0,0.1) [minimum width=0.75cm] (input1) {Input};
\node at (-2.5,-0.1) [minimum width=0.75cm] (input2) {\begin{tabular}{c} Skip\\ Connection\end{tabular}};
\node at (0.0,1.5) [shape=rectangle,minimum height=1.0cm,minimum width=3.8cm,draw] (cnn0) {\begin{tabular}{c} CNN Block\\ $\left(1\times1,C_{i},C_{i}/2\right)$ \end{tabular}};
\node at (0.0,3.2) [shape=rectangle,minimum height=1.0cm,minimum width=3.8cm,draw] (us) {\begin{tabular}{c} Upsample \\ $2\times2$ \end{tabular}};
\draw [->] (input1.north) -- (cnn0.south);
\node at (0.0,4.9) [shape=rectangle,minimum height=1.0cm,minimum width=3.8cm,draw] (con) {Concatenate};
\node at (0.0,6.6) [shape=rectangle,minimum height=1.0cm,minimum width=3.8cm,draw] (cnn1) {\begin{tabular}{c} CNN Block\\ $\left(3\times3,C_{i},C_{i}/2\right)$ \end{tabular}};
\node at (0.0,8.3) [shape=rectangle,minimum height=1.0cm,minimum width=3.8cm,draw] (cnn2) {\begin{tabular}{c} CNN Block\\ $\left(3\times3,C_{i}/2,C_{i}/2\right)$ \end{tabular}};
\node at (0.0,9.6) [minimum width=0.75cm] (output) {Output};
\draw [->] (cnn0.north) -- (us.south);
\draw [->] (us.north) -- (con.south);
\draw [->] (con.north) -- (cnn1.south);
\draw [->] (cnn1.north) -- (cnn2.south);
\draw [-] (input2.north) -- ($ (input2.north) + (0.0,4.45)$);
\draw [->] (cnn2.north) -- (output.south);
\draw [->] ($ (input2.north) + (0.0,4.45)$) -- (con.west);
\node at (2.2,8.3) [minimum width=0.75cm] (times) {$\times2$};
 }
 }
 \caption{Overview of the Expansion Block: For a general Input shape of $T_{i}\times K_{i}\times C_{i}$, the Skip Connection and Output have shape $2T_{i}\times 2K_{i}\times C_{i}/2$.}
 \label{fig:expansion}
\end{figure}

 
The mask estimation network also includes an expansion path which applies a series of \textit{Expansion Blocks}, defined in Figure \ref{fig:expansion}, in order to reconstruct the original resolution of the feature map. Within each \textit{Expansion Block}, a \textit{CNN Block} is first applied to halve the number of channels, followed by nearest-neighbor upsampling with a $2\times2$ window. The feature map is then concatenated with the skip connection from the corresponding level in the contraction path, doubling the number of channels. A \textit{CNN Block} is then applied to halve the number of channels in the feature map, followed by two additional \textit{CNN Blocks}. After the expansion path, a \textit{CNN Block} with Sigmoid activation functions is applied to generate a mask tensor $\m$ which is of the same shape as $\z$ and contains output values in the range $\left[0,1\right]$ which are appropriate for multiplicative masking.
\par
\begin{figure}
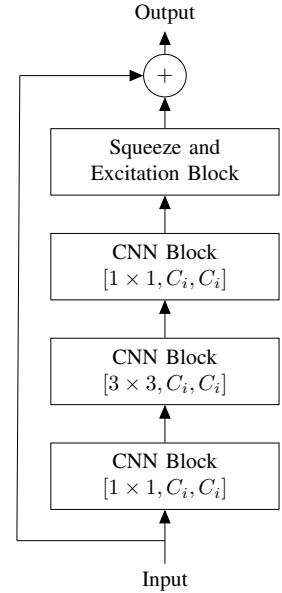

\centering
\scalebox{0.82}{
  \tikz{
\node at (0.0,0.0) [minimum width=0.75cm] (input) {Input};
\node at (0.0,1.7) [shape=rectangle,minimum height=1.0cm,minimum width=3.7cm,draw] (cnn1) {\begin{tabular}{c} CNN Block\\ $\left[1\times1,C_{i},C_{i}\right]$ \end{tabular}};
\node at (0.0,3.4) [shape=rectangle,minimum height=1.0cm,minimum width=3.7cm,draw] (cnn2) {\begin{tabular}{c} CNN Block\\ $\left[3\times3,C_{i},C_{i}\right]$ \end{tabular}};
\node at (0.0,5.1) [shape=rectangle,minimum height=1.0cm,minimum width=3.7cm,draw] (cnn3) {\begin{tabular}{c} CNN Block\\ $\left[1\times1,C_{i},C_{i}\right]$ \end{tabular}};
\node at (0.0,6.8) [shape=rectangle,minimum height=1.0cm,minimum width=3.7cm,draw] (se) {\begin{tabular}{c} Squeeze and \\ Excitation Block \end{tabular}};
\node at (0.0,8.2) [shape=circle,minimum height=0.5cm,minimum width=0.5cm,draw] (sum) {$+$};
\node at (0.0,9.2) [minimum width=0.75cm] (output) {Output};
\draw [->] (input.north) -- (cnn1.south);
\draw [->] (cnn1.north) -- (cnn2.south);
\draw [->] (cnn2.north) -- (cnn3.south);
\draw [->] (cnn3.north) -- (se.south);
\draw [->] (se.north) -- (sum.south);
\draw [->] (sum.north) -- (output.south);
\draw [-] ($ (cnn1.south) + (0.0,-0.5) $) -- ($ (cnn1.south) + (-2.4,-0.5) $);
\draw [-] ($ (cnn1.south) + (-2.4,-0.5) $) -- ($ (sum.west) + (-2.0,0.0) $);
\draw [->] ($ (sum.west) + (-2.0,0.0) $) -- (sum.west);
 }
 }
 \caption{Overview of the Squeeze-and-Excitation Residual Network (SE-ResNet) Block: The Input and Output have the same shape, and $C_{i}$ denotes the number of channels in the Input feature map.}
 \label{fig:senet}
\end{figure}

 
As Figure \ref{fig:mask} illustrates, the modeling capacity of the mask estimation network can be improved with the inclusion of base-level processing between the U-Net's contraction and expansion paths. 
The example architecture here utilizes a base-level network consisting of  $5$ Squeeze-and-Excitation Residual Network (SE-ResNet) blocks \cite{Hu2018}.  These blocks, detailed in Figure \ref{fig:senet}, combine Squeeze-and-Excitation (SE) channel calibration with a Residual Network (ResNet) topology \cite{He2016}, and have been shown to be effective in a variety of tasks.
\section{The MSAE-UNet Training Process}
\label{sec:training}
This section presents the training process of the MSAE-UNet. Specifically, the derivation of target signals is discussed. Additionally, the generation of parallel training data is outlined. Finally, the training loss is described.
\subsection{Designing Target Signals}
\label{sec:target_signals}
Typically, end-to-end systems are trained with parallel data in which known clean speech is corrupted with additive noise; the system learns the inverse mapping from $\x$ to $\s$ through minimization of some distance measure $d\left(\s,\sh\right)$. However, in many realistic environments, speech signals are captured in the presence of reverberation in addition to additive noise and other distortions. While recent end-to-end speech enhancement systems have been shown successful at suppressing additive noise \cite{Fu2018,Pandey2018,Macartney2018,Germain2018,Giri2019,Kolbaek2020}, few studies have successfully addressed suppression of reverberation with an end-to-end system \cite{Luo2018,Borgstrom2021}. Training an end-to-end enhancement system to learn the mapping from $\x$ to $\s$ in the presence of reverberation may be difficult due to the phase distortion caused by convolutional noise.
\par
In this study we utilize the target signal design proposed in \cite{Borgstrom2021}, which enables the enhancement system to learn joint suppression of additive and convolutional distortion. Let $\vv$ be the reverberated-only version of $\s$, which excludes any further distortion present in $\x$. The STFT of $\s$ is given by $\Ss_{m,l}$ where $m$ and $l$ are the frequency channel and frame indices. Let $\V_{m,l}$ be defined similarly. An enhanced version of $\V$ can be obtained using an oracle Wiener Filter,
\begin{equation}
\label{eqn:oracle_filter}
\Ss^{*}_{m,l}=\max\left\{\min\left\{\frac{\left|\Ss_{m,l}\right|^{2}}{\left|\V_{m,l}\right|^{2}},1\right\},0\right\}\V_{m,l}.
\end{equation}
The corresponding waveform, $\s^{*}$, can be synthesized via the inverse STFT. The signal $\s^{*}$ then represents a version of the reverberant signal $\vv$ with the majority of late reflections suppressed, but with the phase distortion introduced by early reflections still present. This allows an end-to-end system to be trained to perform joint suppression of noise and reverberation by learning a mapping from $\x$ to $\s^{*}$ through the minimization of some distance measure $d\left(\s^{*},\sh\right)$.
\subsection{Generating Training Data}
\label{sec:training_data}
Training an end-to-end system requires parallel data consisting of the original (clean) signal, $\s$, and its corrupted, observed counterpart, $\x$. Utilizing the target signals discussed in Section \ref{sec:target_signals} also requires the reverberated-only version, $\vv$. In practice, field recorded data of this nature is relatively limited and not of sufficient quantity or scope to be useful for (at least initial) training purposes. As a result, considerable care has been taken to synthetically generate realistic parallel data representing a broad range of operational conditions.
\par 
Our in-house training material was dynamically generated from an extensive corpus of speech sources, reverberant channels, and noise conditions. The clean speech consists of 456 hours of audio collected from multiple publicly available corpora (e.g. the Linguistic Data Consortium (LDC), LibriSpeech\cite{librispeech}, TIMIT\cite{timit}). Currently 2458 unique talkers are utilized, representing a spectrum of languages and ages. Clean speech signals, $\s$, are randomly selected from this set. The generator then applies a room impulse response to produce $\vv$, and has the capacity to incorporate varying room impulse responses (currently 1708) derived from a number of sources (including \cite{air,voice-home,Voices}) to recreate a generalized ensemble of enclosure sizes and types. Additive noise is then added to produce the observed signal $\x$. The noise conditions are delineated as background (i.e. ambient) and non-stationary.  The 250+ hours of noise samples (collected primarily from the MUSAN\cite{musan}, Voice-Home\cite{voice-home}, Noisex\cite{noisex}, and AudiSet\cite{timit}), represent 218 distinct background and 88756 non-stationary noise conditions. In an effort to further expand the scope of observed material, the room impulse responses and additive noise signals are perturbed by random rate-resampling and spectral-shaping. Additionally, the resulting mixed signal is subjected to multiple potential modifications associated with the signal acquisition process, such as clipping, quantization, and the presence of interceding communications channel.      
\subsection{Training Loss}
\label{sec:pmse}
Training an end-to-end speech enhancement system requires a distance measure which operates on time-domain samples. Initial studies on end-to-end enhancement systems optimized network parameters to minimize the mean square-error (MSE) between the output waveform, $\sh$, and the clean waveform, $\s$,
\begin{align}
\label{eqn:mmse}
d_{MSE}\left(\s,\sh\right)=\frac{1}{\D}\sum^{\D}_{n=1}&\biggl(s\left(n\right)-\shh\left(n\right)\biggr)^{2}.
\end{align}
However, the MSE does not take into account properties of human perception of speech, and may not result in an enhanced signal which optimizes perceptual quality. Recent studies have proposed loss functions which address these issues \cite{Pandey2018,Zhao2018,Fu2018,Fu2019,Xu2019,Kolbaek2020}. In this study, however, we leverage the perceptually-motivated MSE (pMSE) distance proposed in \cite{Borgstrom2021}.
\par
The MSE loss from (\ref{eqn:mmse}) can be generalized to include the effects of both signal pre-emphasis \cite{Quatieri2008} and $\mu$-law amplitude companding  \cite{Jayant1974}, leading to
\begin{align}
\label{eqn:perc_mmse}
d_{pMSE}&\left(\s,\sh\right)=\\\nonumber
&\frac{1}{\D}\sum^{\D}_{n=1}\biggl(\fmu\left(s\left(n\right)-\preemp s\left(n-1\right)\right)\biggr.\\\nonumber
&\ \ \ \ \ \ \ \ \ \ \ \ \biggl.-\fmu\left(\shh\left(n\right)-\preemp\shh\left(n-1\right)\right)\biggr)^{2},
\end{align}
where $\preemp$ is the pre-emphasis constant, and the $\mu$-law companding function is given by
\begin{align}
\fmu\bigl(s\left(n\right)\bigr)=\textrm{sign}\bigl(s\left(n\right)\bigr)\frac{\log\left(1+\mupar\left|s\left(n\right)\right|\right)}{\log\left(1+\mupar\right)}.
\end{align}
Equation (\ref{eqn:perc_mmse}) offers a generalized distance measure which can be tuned to account for various properties of human perception via the parameters $\preemp$ and $\mupar$. Note that for settings $\preemp=0.0$ and $\mupar\rightarrow0.0$, the proposed measure is equivalent to the standard MSE in (\ref{eqn:mmse}). More details regarding the motivation for the pMSE cost can be found in \cite{Borgstrom2021}.
\par 
Note that several studies have proposed the use of spectral-based cost functions which don't take into account the phase signal \cite{wang2018,Kolbaek2019,Kolbaek2020,defossez2020}. 
During our experimentation, we found that while the inclusion of such spectral-based components in the overall training loss provided some advantage with the objective speech quality measures, they also lead to the inclusion of 'buzzy' artifacts in the resulting speech, particularly in the higher frequencies. For this reason the time-domain pMSE loss, which offers both a degree of perceptual modelling as well as a sensitivity to speech phase distortion, was utilized exclusively in this study.
\par
In \cite{Borgstrom2021}, a dual-path loss was proposed for mask-based end-to-end enhancement systems in order to ensure that the autencoder path provides high quality reconstruction when encoder and decoder filters are trained. The dual-path loss is given by $d_{pMSE}\left(\s,\sh\right)+d_{pMSE}\left(\x,\xh\right)$, where the additional second term enforces the approximation from (\ref{eqn:autoencoder_approx}). In this study, the dual-path loss was utilized whenever MSAE kernels were learnable, as discussed in Section \ref{sec:learned_filters}. 
\par
The construction of training data can have significant impact on the behavior of the resulting neural network enhancement systems. For example, the ratio of active speech to inactive speech within the data can change the system's emphasis on speech quality versus noise suppression. To better control this tradeoff during training, a prior probability of active speech can be induced by weighting individual training waveforms. It is straightforward to determine whether a clean reference sample contains a significant amount of active speech. For a given training batch, let $M_{1}$ and $M_{0}$ denote the total number of samples that are determined to correspond to active speech and inactive speech, respectively. To introduce an effective prior, $\prior\in\left[0,1\right]$, active speech and inactive speech segments can be weighted by $\prior\left(M_{0}+M_{1}\right)/M_{1}$ and $\left(1-\prior\right)\left(M_{0}+M_{1}\right)/M_{0}$, respectively. In this way, using a large $\prior$ focuses network training on speech quality, whereas a small value puts more emphasis on noise suppression.
\section{Experimental Results}
\label{sec:results}
This section presents experimental results for speech enhancement using MSAE-Unet, both in terms of objective speech quality metrics and automatic speech recognition performance. During experimentation, a variety of MSAE configurations are analyzed. For all systems, input waveforms are split into windows of dimension $\D=20480$, corresponding to $1.28$ sec, with $50\%$ overlap, before processing. Output waveforms are then reconstructed using overlap-and-add. All windows are scaled to unit variance prior to enhancement, and gains are reintroduced after network processing. Networks are trained using target signals generated according to Section \ref{sec:target_signals}. Parallel training data is generated as discussed in Section \ref{sec:training_data}. Networks are optimized by minimizing the pMSE loss described in Section \ref{sec:pmse}, using an active speech prior of $\prior=0.75$. In all testing, the minimum gain, $\gmin$, is set to $-50$dB, which is observed to provide a good trade-off between noise suppression and speech quality during informal listening.
\subsection{Objective Speech Quality}
Speech enhancement experiments are performed on the Voice Cloning Toolkit (VCTK) Noisy and Reverberant Corpus \cite{vctk}, a parallel clean-corrupted set comprised of $824$ speech files with synthetically added reverberation and background noise. To assess performance, a variety of objective speech quality metrics are used. Specifically, results are reported in terms of Perceptual Evaluation of Speech Quality (PESQ) \cite{pesq} and Short-time Objective Intelligibility (STOI) \cite{taal2010}. 
Results are also reported in terms of the Composite Signal (CSIG), Composite Background (CBAK), and Composite Overall (COVL) quality scores from \cite{hu2007}, which are weighted combinations of various signal quality measures where the weightings are trained to correlate with subjective listening tests. Finally, the relative improvement in COVL compared to the unprocessed signal is provided to offer more clarity. In all tables, bold entries denote the best result for each metric. 
\subsubsection{The Effect of Analysis Window Duration}
The first set of experiments aim at exploring the effect of analysis window duration in neural network speech enhancement systems. Table \ref{tab:results_conventional} provides speech quality measures on the VCTK Noisy and Reverberant Corpus for MSAE-UNet with the single-branch MSAE configuration, $\left(1,-,T_{o}\right)$, and varying $T_{o}$. Note that short window durations, e.g. $T_{o}=5$ms, correspond to conventional wideband analysis in the autoencoder, whereas long windows, e.g. $T_{o}=40$ms, correspond to conventional narrowband analysis. From Table \ref{tab:results_conventional}, it can be observed that analysis window duration has a significant impact on speech quality metrics, with the best performance achieved for intermediate durations.
With respect to COVL, an analysis window duration of $T_{o}=10$ms maximizes the quality of enhanced speech. These results show the sensitivity of neural network speech enhancement to the design of analysis windows, and motivates the MSAE framework which can represent separate frequency bands with appropriate window durations.
\begin{table*}[th]
	\caption{The Effect of Analysis Window Duration in End-to-end Neural Network Speech Enhancement: Results on the VCTK Noisy and Reverberant Corpus are reported for mask-based enhancement systems with single-branch autoencoders, corresponding to MSAE frameworks with $\left(1,-,T_{o}\right)$ configurations and varying analysis window durations $\To$.}
  \label{tab:results_conventional}
  \centering
  \begin{tabular}{c c c | c c c c c | c}
	\hline
	\multicolumn{3}{c|}{MSAE Confirguration} & \multirow{2}{*}{PESQ} & \multirow{2}{*}{STOI} & \multirow{2}{*}{CSIG} & \multirow{2}{*}{CBAK} & \multirow{2}{*}{COVL} & \multirow{2}{*}{$\Delta$COVL $\left(\%\right)$}\\
	$B$ & $\Q$ & $\To$ (ms) & & & & & &\\\hline\hline
    \multirow{6}{*}{$1$} & \multirow{6}{*}{$-$} & $2.5$ & $1.586$ & $84.24$ & $2.327$ & $1.884$ & $1.868$ & $44.4$ \\
    & & $5$ & $1.609$ & $\mathbf{85.22}$ & $2.346$ & $1.921$ & $1.893$ & $46.3$ \\
    & & $10$ & $1.637$ & $85.17$ & $\mathbf{2.355}$ & $1.961$ & $\mathbf{1.920}$ & $\mathbf{48.4}$ \\
    & & $20$ & $\mathbf{1.639}$ & $85.01$ & $2.308$ & $\mathbf{1.978}$ & $1.904$ & $47.1$ \\
    & & $40$ & $1.581$ & $83.99$ & $2.100$ & $1.902$ & $1.757$ & $35.8$ \\
    & & $80$ & $1.527$ & $81.19$ & $1.903$ & $1.832$ & $1.618$ & $25.0$ \\
  \hline
  \end{tabular}
\end{table*}
\subsubsection{The Multi-Branch MSAE Framework}
The next set of experiments aim at assessing the benefits of the proposed MSAE framework within mask-based enhancement systems. An important parameter of the MSAE is the number of branches, $B$. Table \ref{tab:results_bands} provides speech quality measures on the VCTK Noisy and Reverberant Corpus for MSAE-UNet with the multi-branch MSAE configuration, $\left(B,1.5,2.5\textrm{ms}\right)$, and varying number of branches. All systems use a default dyadic spectral band design, corresponding to the quality factor $Q=1.5$. It can be observed that speech quality improves significantly for MSAE configurations with multiple branches, when compared to the single-branch system, with the best performance achieved for $B=5$.
Specifically, the $B=5$ MSAE provides an additional $16\%$ relative improvement in COVL compared to the single-branch autoencoder. These results clearly show the benefit of the proposed multiscale autoencoder, relative to conventional systems.
\begin{table*}[th]
	\caption{The Effect of Using Multiple Branches in the MSAE Framework: Results on the VCTK Noisy and Reverberant Corpus are reported for multi-branch MSAE configurations, $\left(B,1.5,2.5\textrm{ms}\right)$, with varying number of branches $B$.}
  \label{tab:results_bands}
  \centering
  \begin{tabular}{c c c | c c c c c | c}
	\hline
	\multicolumn{3}{c|}{MSAE Confirguration} & \multirow{2}{*}{PESQ} & \multirow{2}{*}{STOI} & \multirow{2}{*}{CSIG} & \multirow{2}{*}{CBAK} & \multirow{2}{*}{COVL} & \multirow{2}{*}{$\Delta$COVL $\left(\%\right)$}\\
	$B$ & $\Q$ & $\To$ (ms) & & & & & &\\\hline\hline
    $1$ &\multirow{7}{*}{$1.5$} & \multirow{7}{*}{$2.5$} & $1.586$ & $84.24$ & $2.327$ & $1.884$ & $1.868$ & $44.4$ \\
    $2$ & & & $1.609$ & $84.50$ & $2.413$ & $1.917$ & $1.928$ & $49.0$\\
    $3$ & & & $1.679$ & $84.93$ & $2.517$	& $1.971$ & $2.024$ & $56.4$ \\
    $4$ & & & $1.740$ & $85.65$	& $2.582$ & $2.034$ & $2.094$ & $61.8$ \\
    $5$ & & & $\mathbf{1.783}$ & $86.10$ & $\mathbf{2.595}$ & $\mathbf{2.051}$ & $\mathbf{2.121}$ & $\mathbf{63.9}$ \\
    $6$ & & & $1.763$ & $86.14$ & $2.493$ & $2.041$ & $2.059$ & $59.1$ \\
    $7$ & & & $1.756$ & $\mathbf{86.50}$ & $2.443$ & $2.034$ & $2.026$ & $56.6$ \\
  \hline
  \end{tabular}
\end{table*}
\subsubsection{Spectral Band Design within the MSAE}
The MSAE is also parameterized by the Constant-Q band design. The next experiments explore the role of the quality factor, $Q$, in enhancement performance. Table \ref{tab:results_q} provides speech quality measures on the VCTK Noisy and Reverberant Corpus for the MSAE configuration $\left(5,Q,2.5\textrm{ms}\right)$, for different values of $Q$. It can be observed that relative to a dyadic band design, speech quality can be improved by tuning the quality factor to $Q=2.0$, which corresponds to narrower bands and a higher degree to frequency warping than that illustrated in Figure \ref{fig:freq_responses}. Specifically, the improved band design provides an additional $2\%$ relative improvement in COVL.
\begin{table*}[th]
	\caption{The Effect of Spectral Band Design in the MSAE Framework: Results on the VCTK Noisy and Reverberant Corpus are reported for various values of $Q$ in the Constant-Q band design.}
  \label{tab:results_q}
  \centering
  \begin{tabular}{c c c | c c c c c | c}
	\hline
	\multicolumn{3}{c|}{MSAE Confirguration}  & \multirow{2}{*}{PESQ} & \multirow{2}{*}{STOI} & \multirow{2}{*}{CSIG} & \multirow{2}{*}{CBAK} & \multirow{2}{*}{COVL} & \multirow{2}{*}{$\Delta$COVL $\left(\%\right)$}\\
	$B$ & $\Q$ & $\To$ (ms) & & & & & & \\\hline\hline
    \multirow{4}{*}{$5$} & $1.0$ & \multirow{4}{*}{$2.5$} & $1.691$ & $85.30$ & $2.459$ & $1.978$ & $2.000$ & $54.6$ \\
    & $1.5$ & & $1.783$ & $86.10$ & $\mathbf{2.595}$ & $2.051$ & $2.121$ & $63.9$ \\
    & $2.0$ & & $\mathbf{1.805}$ & $\mathbf{86.62}$ & $2.591$ & $\mathbf{2.071}$ & $\mathbf{2.131}$ & $\mathbf{64.7}$ \\
    & $2.5$ & & $1.790$	& $86.54$ & $2.553$ & $2.058$ & $2.103$ & $62.5$ \\
  \hline
  \end{tabular}
\end{table*}
\subsubsection{Trainable MSAE Filters}
The final set of experiments explore the use of trainable encoder and decoder filters within the MSAE framework. As discussed in Section \ref{sec:learned_filters}, the MSAE encoder and decoder networks are composed of differentiable operations, and can be constructed entirely using standard neural network blocks, allowing the kernels $\W$ to be trained jointly with the mask estimation network. Table \ref{tab:results_train} compares speech enhancement performance of the MSAE-UNet with the $\left(5,2.0,2.5\textrm{ms}\right)$ MSAE configuration, with fixed and learned kernels. In the latter case, the overcompleteness factor, $\overcomp$, is specified. In the table, it can be observed that allowing MSAE filters to adapt during training provides performance improvements across speech quality metrics. Specifically, the use of learned filters and an overcompleteness factor $\overcomp=1.5$ provides an additional $6\%$ relative improvement in COVL. Note that in the remainder of this section, only results for this MSAE-UNet configuration will be reported.
\begin{table*}[th]
	\caption{The Effect of Learnable Encoder and Decoder Filters in the MSAE Framework: Results on the VCTK Noisy and Reverberant Corpus are reported for the MSAE configuration $\left(5,2.0,2.5\textrm{ms}\right)$, where the kernels $\W$ are either fixed according to (\ref{eqn:W}), or adapted during network training. In the case of adaptation, the overcompleteness factor, $\overcomp$, is specified.}
  \label{tab:results_train}
  \centering
  \begin{tabular}{c c c | c c | c c c c c | c}
	\hline
	\multicolumn{3}{c|}{MSAE Confirguration} & \multicolumn{2}{c|}{$\W$} & \multirow{2}{*}{PESQ} & \multirow{2}{*}{STOI} & \multirow{2}{*}{CSIG} & \multirow{2}{*}{CBAK} & \multirow{2}{*}{COVL} & \multirow{2}{*}{$\Delta$COVL $\left(\%\right)$} \\
	$B$ & $\Q$ & $\To$ (ms) & Parameters & $\overcomp$ & & & & & \\\hline\hline
    \multirow{4}{*}{$5$} & \multirow{4}{*}{$2.0$} & \multirow{4}{*}{$2.5$} & Fixed & $-$ & $1.805$ & $86.62$ & $2.591$ & $2.071$ & $2.131$ & $64.7$ \\\cline{4-11}
    & & & \multirow{3}{*}{Adapted} & $1.0$ & $1.815$ & $86.28$ & $2.666$ & $2.064$ & $2.180$ & $68.5$ \\
    & & & & $1.5$ & $\mathbf{1.834}$ & $87.07$ & $\mathbf{2.700}$ & $\mathbf{2.074}$ & $\mathbf{2.206}$ & $\mathbf{70.5}$ \\
    & & & & $2.0$ & $1.816$ & $\mathbf{87.24}$ & $2.690$ & $2.071$ & $2.192$ & $69.4$ \\
  \hline
  \end{tabular}
\end{table*}
\subsubsection{An Ablation Study}
In order to give a clear summary of the previous experiments, Table \ref{tab:ablation} provides an ablation study of neural network speech enhancement using the MSAE framework. The table first includes objective speech quality metrics for the original degraded signals in the VCTK Noisy and Reverberant Corpus. The next row provides results for a baseline mask-based enhancement system which uses a single-branch autoencoder with a $10$ms anaylsis window, and is trained to learn the mapping from noisy signals $\x$ to clean signals $\s$, using the standard MSE loss. Each successive row then provides results when cumulatively adding an additional feature to the enhancement system.
\par
In Table \ref{tab:ablation}, the poor results of the baseline system may be due to the use of clean target signals. The presence of potentially severe reverberation in the training data may introduce a high degree of phase distortion, making it difficult for an end-to-end system to learn the mapping from the noisy signal $\x$ to the clean signal $\s$. In the third row, Wiener-filtered target signals described in Section \ref{sec:target_signals} are instead used during training, leading to a sigificant improvement in enhancement performance. In the fourth row, the conventional MSE loss is replaced by the perceptual MSE during training, as discussed in Section \ref{sec:pmse}, leading to further performance improvements.
\par
In Table \ref{tab:ablation}, the fifth row provides results for a multi-branch MSAE. Specifically, the $\left(5,1.5,2.5\textrm{ms}\right)$ MSAE configuration with fixed $\W$ filters is used, providing further performance improvements. Next, the Constant-Q spectral band design within the MSAE was tuned from a dyadic decomposition to a quality factor of $Q=2.0$, yielding the results in the sixth row of the table. Finally, the seventh row provides results for the use of adapted MSAE kernels, providing a slight performance improvement in certain objective metrics.
\begin{table*}[th]
	\caption{An Ablation Study of End-to-end Mask-Based Systems with the MSAE Framework: The first row provides quality measures for the original signals in the VCTK Noisy and Reverberant Corpus. Each successive row provides results when cumulatively adding an additional feature to the enhancement system.}
  \label{tab:ablation}
  \centering
  \begin{tabular}{l | c c c c c | c}
	\hline
	System & PESQ & STOI & CSIG & CBAK & COVL & $\Delta$COVL $\left(\%\right)$\\\hline\hline
	Original & $1.218$ & $63.79$ & $1.553$ & $1.357$ & $1.294$ & $0.0$ \\\hline
    Baseline System & $1.433$ & $78.84$ & $1.465$ & $1.755$ & $1.351$ & $4.4$\\
    Wiener Filtered Target Signals (Section \ref{sec:target_signals}) & $1.517$ & $83.45$ & $1.997$ & $1.815$ & $1.675$ & $29.4$ \\
    Perceptual MSE Training Loss (Section \ref{sec:pmse}) & $1.613$ & $84.86$ & $2.414$ & $1.931$ & $1.934$ & $49.5$ \\
    Multiscale Autoencoder (Section \ref{sec:encoder}, \ref{sec:decoder}) & $1.783$ & $86.10$ & $2.595$ & $2.051$ & $2.121$ & $63.9$ \\
    Spectral Band Design (Section \ref{sec:quality_factor}) & $1.805$ & $86.62$ & $2.591$ & $2.071$ & $2.131$ & $64.7$ \\
    Learned Autoencoder Filters (Section \ref{sec:learned_filters}) & $\mathbf{1.834}$ & $\mathbf{87.07}$ & $\mathbf{2.700}$ & $\mathbf{2.074}$ & $\mathbf{2.206}$ & $\mathbf{70.5}$ \\
  \hline
  \end{tabular}
\end{table*}
\subsubsection{Comparison to State-of-the-Art Enhancement Systems}
Finally, the MSAE-Unet was compared to state-of-the-art neural network speech enhancement systems, and a variety of baselines was chosen to assess the performance of the proposed system. First, the MetricGAN+ \cite{fu2021} system uses a Bidirectional Long Short Term Memory (BLSTM) architecture \cite{weninger2015}, trained within a Generative Adversarial Network (GAN) \cite{goodfellow2020} framework using a perceptually-motivated loss. Next, the Mimic system \cite{bagchi2018} uses a neural network architecture, and is trained jointly to both minimize the MSE loss and to improve senone classification of the enhanced speech. Finally, the DEMUCS system \cite{defossez2020} uses a U-Net architecture, and is trained with a composite time-domain and spectral-domain loss. Note that the MetricGAN+ and Mimic models are provided by the SpeechBrain Toolkit \cite{speechbrain}, and were trained with data from the VoiceBank \cite{VoiceBank} and Deep Noise Suppression (DNS) \cite{reddy2020} corpora. The DEMUCS model was trained on the VCTK Noisy \cite{valentini2016} and DNS corpora. Table \ref{tab:baselines} provides a performance comparison of MSAE-UNet with the baseline systems on the VCTK Noisy and Reverberant Corpus. Additionally, it includes the number of trainable parameters in each model studied. As can be observed, the proposed system achieves significant performance improvements across speech quality metrics. Specifically, the MSAE-UNet provides additional relative improvements in COVL of $12\%$-$24\%$ compared to the baseline systems. This is especially noteworthy considering the small size of MSAE-UNet relative to the Mimic and DEMUCS models.
\begin{table*}[th]
	\caption{Speech Enhancement Performance on the VCTK Noisy and Reverberant Corpus}
  \label{tab:baselines}
  \centering
  \begin{tabular}{l | c | c c c c c | c}
	\hline
	System & $\#$Params & PESQ & STOI & CSIG & CBAK & COVL & $\Delta$COVL $\left(\%\right)$\\\hline\hline
	Original & $-$ & $1.218$ & $63.79$ & $1.553$ & $1.357$ & $1.294$ & $0.0$ \\\hline
	MetricGAN$+$ \cite{fu2021} & $1.9$ M& $1.807$ & $79.05$ & $2.219$ & $1.813$ & $1.895$ & $46.4$ \\
	Mimic \cite{bagchi2018} & $22.3$ M & $1.564$ & $81.69$ & $2.660$ & $1.900$ & $2.045$ & $58.0$ \\
	DEMUCS \cite{defossez2020} & $60.3$ M & $1.634$ & $84.46$ & $2.588$ & $1.964$ & $2.055$ & $58.8$ \\
    MSAE-UNet & $6.7$ M & $\mathbf{1.834}$ & $\mathbf{87.07}$ & $\mathbf{2.700}$ & $\mathbf{2.074}$ & $\mathbf{2.206}$ & $\mathbf{70.5}$ \\
  \hline
  \end{tabular}
\end{table*}
\subsection{Automatic Speech Recognition}
Enhancement can not only improve the perceptual quality of distorted speech, alleviating listener fatigue, but can also improve performance of applications such as automated speech recognition (ASR). This section provides experimentation using enhancement as pre-processing for speech recognition. In order to provide general results, two ASR models are studied, namely Deep Speech \cite{hannun2014} and the \textit{Large} variant of Whisper \cite{whisper}, both representing state-of-the-art end-to-end neural network approaches. Experiments are performed on the VCTK Noisy and Reverberant Corpus and the Voices Obscured in Complex Environmental Settings (VOiCES) Corpus \cite{Voices}. The latter is a large set of far-field audio collected by replaying speech and noise distractor signals within various reverberant rooms. The corpus captures a wide range of acoustic environments by varying room dimensions, noise types, microphone types, and microphone placements. Throughout experimentation, results are reported as Word Error Rate (WER), calculated using the NIST Scoring Toolkit (SCTK) \cite{sctk}.
\par 
Table \ref{tab:results_asr} provides ASR results for the Deep Speech and Whisper models, when using enhancement as a pre-processing step. It can be observed that with the Deep Speech model, the baseline enhancement systems generally result in performance degradation relative to the original signals. This degradation is especially large for the VOiCES Corpus. The MSAE-UNet system, however, provides ASR improvements for both benchmarks. For the Whisper model, the baseline enhancement systems again lead to performance degradation in many cases. The MSAE-UNet system, however, provides clear performance improvements for the VCTK benchmark, and minor degradation for the VOiCES Corpus. 
\begin{table*}[th]
	\caption{ASR Performance on the VOiCES and VCTK Noisy and Reverberant Corpora: Results are reported for the Whisper (Large) and the Deep Speech models, in terms of Word Error Rate $\left(\%\right)$.}
  \label{tab:results_asr}
  \centering
  \begin{tabular}{l | c c | c c }
    \hline
	\multirow{2}{*}{System} & \multicolumn{2}{c|}{Deep Speech} & \multicolumn{2}{c}{Whisper} \\
	& VCTK & VOiCES & VCTK & VOiCES  \\\hline\hline
    Original & $50.2$& $46.5$ & $31.3$ & $\mathbf{26.0}$\\\hline
    MetricGAN$+$ \cite{fu2021} & $60.0$& $76.7$ & $36.9$ & $52.6$ \\
    Mimic \cite{bagchi2018} & $53.8$& $86.9$ & $34.6$ & $72.9$ \\
    DEMUCS \cite{defossez2020} & $47.9$& $60.9$ & $31.6$ & $52.2$ \\
    MSAE-UNet & $\mathbf{43.6}$ & $\mathbf{37.0}$ & $\mathbf{26.3}$ & $29.9$ \\
  \hline
  \end{tabular}
\end{table*}
\section{Conclusion}
\label{sec:conclusion}
This paper proposed the multiscale autoencoder (MSAE) for mask-based end-to-end neural network speech enhancement. This framework provides the encoder and decoder mappings for such networks, and is not specific to the mask estimation architecture. The MSAE performs spectral decomposition of an input waveform within separate band-limited branches, each operating with a different rate and scale, to extract a sequence of multiscale embeddings. The proposed framework features intuitive parameterization of the autoencoder, including a flexible spectral band design based on the Constant-Q transform. Additionally, the MSAE is constructed entirely of differentiable operators, allowing it to be implemented within an end-to-end neural network, and discriminatively trained for the task of speech enhancement. To assess the performance of the proposed MSAE framework, it was integrated with an example mask estimator based on the $U$-Net architecture. The resulting end-to-end enhancement system was shown to outperform several state-of-the-art speech enhancement methods both in terms of objective speech quality metrics and automatic speech recognition accuracy.
\bibliographystyle{IEEEbib}
\bibliography{mybib}

\begin{thebibliography}{10}

\bibitem{loizou2010}
P.~C. Loizou and G.~Kim,
\newblock ``Reasons why current speech-enhancement algorithms do not improve
  speech intelligibility and suggested solutions,''
\newblock {\em IEEE Transactions on Audio, Speech, and Language Processing},
  vol. 19, no. 1, pp. 47--56, 2010.

\bibitem{zekveld2011}
A.~A. Zekveld, S.~E. Kramer, and J.~M. Festen,
\newblock ``Cognitive load during speech perception in noise: The influence of
  age, hearing loss, and cognition on the pupil response,''
\newblock {\em Ear and hearing}, vol. 32, no. 4, pp. 498--510, 2011.

\bibitem{borgstrom2012}
B.~J. Borgstr{\"o}m and A.~McCree,
\newblock ``The linear prediction inverse modulation transfer function
  ({LP-IMTF}) filter for spectral enhancement, with applications to speaker
  recognition,''
\newblock in {\em ICASSP}, 2012, pp. 4065--4068.

\bibitem{virtanen2012}
T.~Virtanen, R.~Singh, and B.~Raj,
\newblock {\em Techniques for noise robustness in automatic speech
  recognition},
\newblock John Wiley \& Sons, 2012.

\bibitem{mandasari2012}
M.~I. Mandasari, M.~McLaren, and D.~A. van Leeuwen,
\newblock ``The effect of noise on modern automatic speaker recognition
  systems,''
\newblock in {\em ICASSP}, 2012, pp. 4249--4252.

\bibitem{li2014}
J.~Li, L.~Deng, Y.~Gong, and R.~Haeb-Umbach,
\newblock ``An overview of noise-robust automatic speech recognition,''
\newblock {\em IEEE/ACM Transactions on Audio, Speech, and Language
  Processing}, vol. 22, no. 4, pp. 745--777, 2014.

\bibitem{sadjadi2014}
S.O. Sadjadi and J.H.L. Hansen,
\newblock ``Blind spectral weighting for robust speaker identification under
  reverberation mismatch,''
\newblock {\em IEEE/ACM transactions on audio, speech, and language
  processing}, vol. 22, no. 5, pp. 937--945, 2014.

\bibitem{Mcaulay1980}
R.~McAulay and M.~Malpass,
\newblock ``Speech enhancement using a soft-decision noise suppression
  filter,''
\newblock {\em IEEE Transactions on Acoustics, Speech, and Signal Processing},
  vol. 28, no. 2, pp. 137--145, 1980.

\bibitem{Ephraim1984}
Y.~Ephraim and D.~Malah,
\newblock ``Speech enhancement using a minimum-mean square error short-time
  spectral amplitude estimator,''
\newblock {\em IEEE Transactions on acoustics, speech, and signal processing},
  vol. 32, no. 6, pp. 1109--1121, 1984.

\bibitem{Ephraim1985}
Y.~Ephraim and D.~Malah,
\newblock ``Speech enhancement using a minimum mean-square error log-spectral
  amplitude estimator,''
\newblock {\em IEEE Transactions on acoustics, speech, and signal processing},
  vol. 33, no. 2, pp. 443--445, 1985.

\bibitem{Cohen2002}
I.~Cohen,
\newblock ``Optimal speech enhancement under signal presence uncertainty using
  log-spectral amplitude estimator,''
\newblock {\em IEEE Signal Processing Letters}, vol. 9, no. 4, pp. 113--116,
  2002.

\bibitem{Borgstrom2021}
B.~J. Borgstr{\"o}m and M.~S. Brandstein,
\newblock ``Speech enhancement via attention masking network ({SEAMNET}): An
  end-to-end system for joint suppression of noise and reverberation,''
\newblock {\em IEEE/ACM Transactions on Audio, Speech, and Language
  Processing}, vol. 29, pp. 515--526, 2021.

\bibitem{weninger2015}
F.~Weninger, H.~Erdogan, S.~Watanabe, E.~Vincent, J.~Le~Roux, J.~R. Hershey,
  and B.~Schuller,
\newblock ``Speech enhancement with {LSTM} recurrent neural networks and its
  application to noise-robust {ASR},''
\newblock in {\em Latent Variable Analysis and Signal Separation}, 2015, pp.
  91--99.

\bibitem{valentini2016}
C.~Valentini-Botinhao, X.~Wang, S.~Takaki, and J.~Yamagishi,
\newblock ``Speech enhancement for a noise-robust text-to-speech synthesis
  system using deep recurrent neural networks.,''
\newblock in {\em Interspeech}, 2016, vol.~8, pp. 352--356.

\bibitem{Zhao2018}
Y.~Zhao, B.~Xu, R.~Giri, and T.~Zhang,
\newblock ``Perceptually guided speech enhancement using deep neural
  networks,''
\newblock in {\em ICASSP}, 2018, pp. 5074--5078.

\bibitem{Pandey2018}
A.~Pandey and D.~Wang,
\newblock ``A new framework for supervised speech enhancement in the time
  domain,''
\newblock in {\em Interspeech}, 2018, pp. 1136--1140.

\bibitem{Macartney2018}
C.~Macartney and T.~Weyde,
\newblock ``Improved speech enhancement with the wave-u-net,''
\newblock {\em arXiv preprint arXiv:1811.11307}, 2018.

\bibitem{Germain2018}
F.~G. Germain, Q.~Chen, and V.~Koltun,
\newblock ``Speech denoising with deep feature losses,''
\newblock {\em Interspeech}, pp. 2723--2727, 2019.

\bibitem{Luo2018}
Y.~Luo and N.~Mesgarani,
\newblock ``Real-time single-channel dereverberation and separation with
  time-domain audio separation network,''
\newblock {\em Proc. Interspeech 2018}, pp. 342--346, 2018.

\bibitem{Soni2018}
M.~H. Soni, N.~Shah, and H.~A. Patil,
\newblock ``Time-frequency masking-based speech enhancement using generative
  adversarial network,''
\newblock in {\em ICASSP}, 2018, pp. 5039--5043.

\bibitem{Fu2018}
S.-W. Fu, T.-W. Wang, Y.~Tsao, X.~Lu, and H.~Kawai,
\newblock ``End-to-end waveform utterance enhancement for direct evaluation
  metrics optimization by fully convolutional neural networks,''
\newblock {\em IEEE/ACM Transactions on Audio, Speech, and Language
  Processing}, vol. 26, no. 9, pp. 1570--1584, 2018.

\bibitem{Koizumi2018}
Y.~Koizumi, N.~Harada, Y.~Haneda, Y.~Hioka, and K.~Kobayashi,
\newblock ``End-to-end sound source enhancement using deep neural network in
  the modified discrete cosine transform domain,''
\newblock in {\em ICASSP}, 2018, pp. 706--710.

\bibitem{bagchi2018}
D.~Bagchi, P.~Plantinga, A.~Stiff, and E.~Fosler-Lussier,
\newblock ``Spectral feature mapping with mimic loss for robust speech
  recognition,''
\newblock in {\em ICASSP}, 2018, pp. 5609--5613.

\bibitem{Pandey2019}
A.~Pandey and D.~Wang,
\newblock ``A new framework for cnn-based speech enhancement in the time
  domain,''
\newblock {\em IEEE/ACM Transactions on Audio, Speech, and Language
  Processing}, vol. 27, no. 7, pp. 1179--1188, 2019.

\bibitem{Fu2019}
Szu-Wei Fu, Chien-Feng Liao, and Yu~Tsao,
\newblock ``Learning with learned loss function: Speech enhancement with
  quality-net to improve perceptual evaluation of speech quality,''
\newblock {\em IEEE Signal Processing Letters}, vol. 27, pp. 26--30, 2019.

\bibitem{Giri2019}
R.~Giri, U.~Isik, and A.~Krishnaswamy,
\newblock ``Attention wave-u-net for speech enhancement,''
\newblock in {\em WASPAA}, 2019, pp. 249--253.

\bibitem{Xu2019}
Z.~Xu, S.~Elshamy, Z.~Zhao, and T.~Fingscheidt,
\newblock ``Components loss for neural networks in mask-based speech
  enhancement,''
\newblock {\em arXiv preprint arXiv:1908.05087}, 2019.

\bibitem{Luo2019}
Y.~Luo and N.~Mesgarani,
\newblock ``Conv-tasnet: surpassing ideal time--frequency magnitude masking for
  speech separation,''
\newblock {\em IEEE/ACM Transactions on Audio, Speech, and Language
  Processing}, vol. 27, no. 8, pp. 1256--1266, 2019.

\bibitem{Casebeer2020}
J.~Casebeer, U.~Isik, S.~Venkataramani, and A.~Krishnaswamy,
\newblock ``Efficient trainable front-ends for neural speech enhancement,''
\newblock in {\em ICASSP}, 2020, pp. 6639--6643.

\bibitem{Xu2020}
Z.~Xu, S.~Elshamy, and T.~Fingscheidt,
\newblock ``Using separate losses for speech and noise in mask-based speech
  enhancement,''
\newblock in {\em ICASSP}, 2020, pp. 7519--7523.

\bibitem{Kolbaek2020}
M.~Kolb{\ae}k, Z.-H. Tan, S.~H. Jensen, and J.~Jensen,
\newblock ``On loss functions for supervised monaural time-domain speech
  enhancement,''
\newblock {\em IEEE/ACM Transactions on Audio, Speech, and Language
  Processing}, vol. 28, pp. 825--838, 2020.

\bibitem{defossez2020}
A.~Defossez, G.~Synnaeve, and Y.~Adi,
\newblock ``Real time speech enhancement in the waveform domain,'' 2020.

\bibitem{Li2021}
A.~Li, W.~Liu, C.~Zheng, C.~Fan, and X.~Li,
\newblock ``Two heads are better than one: A two-stage complex spectral mapping
  approach for monaural speech enhancement,''
\newblock {\em IEEE/ACM Transactions on Audio, Speech, and Language
  Processing}, vol. 29, pp. 1829--1843, 2021.

\bibitem{fu2021}
S.-W. Fu, C.~Yu, T.-A. Hsieh, P.~Plantinga, M.~Ravanelli, X.~Lu, and Y.~Tsao,
\newblock ``Metricgan+: An improved version of metricgan for speech
  enhancement,'' 2021.

\bibitem{subakan2021}
C.~Subakan, M.~Ravanelli, S.~Cornell, M.~Bronzi, and J.~Zhong,
\newblock ``Attention is all you need in speech separation,'' 2021.

\bibitem{zhang2022}
Q.~Zhang, X.~Qian, Z.~Ni, A.~Nicolson, E.~Ambikairajah, and H.~Li,
\newblock ``A time-frequency attention module for neural speech enhancement,''
\newblock {\em IEEE/ACM Transactions on Audio, Speech, and Language
  Processing}, vol. 31, pp. 462--475, 2022.

\bibitem{Williamson2015}
D.~S. Williamson, Y.~Wang, and D.~Wang,
\newblock ``Complex ratio masking for monaural speech separation,''
\newblock {\em IEEE/ACM transactions on audio, speech, and language
  processing}, vol. 24, no. 3, pp. 483--492, 2015.

\bibitem{Tan2019}
K.~Tan and D.~Wang,
\newblock ``Learning complex spectral mapping with gated convolutional
  recurrent networks for monaural speech enhancement,''
\newblock {\em IEEE/ACM Transactions on Audio, Speech, and Language
  Processing}, vol. 28, pp. 380--390, 2019.

\bibitem{Li2021b}
Q.~Li, F.~Gao, H.~Guan, and K.~Ma,
\newblock ``Real-time monaural speech enhancement with short-time discrete
  cosine transform,'' 2021.

\bibitem{olshausen2002}
B.~Olshausen and K.~O'Connor,
\newblock ``A new window on sound,''
\newblock {\em Nature Neuroscience}, vol. 5, pp. 292--4, 05 2002.

\bibitem{oppenheim1970speech}
A.~V. Oppenheim,
\newblock ``Speech spectrograms using the fast fourier transform,''
\newblock {\em IEEE spectrum}, vol. 7, no. 8, pp. 57--62, 1970.

\bibitem{smith2011}
J.~O. Smith,
\newblock {\em Spectral Audio Signal Processing},
\newblock 2011,
\newblock online book, 2011 edition.

\bibitem{brown1991}
J.~C. Brown,
\newblock ``Calculation of a constant {Q} spectral transform,''
\newblock {\em The Journal of the Acoustical Society of America}, vol. 89, pp.
  425--434, 1991.

\bibitem{mallat2008wavelet}
S.~Mallat,
\newblock {\em A Wavelet Tour of Signal Processing: The Sparse Way},
\newblock Elsevier Science, 2008.

\bibitem{Xie2017}
S.~Xie, R.~Girshick, P.~Doll{\'a}r, Z.~Tu, and K.~He,
\newblock ``Aggregated residual transformations for deep neural networks,''
\newblock in {\em CVPR}, 2017, pp. 1492--1500.

\bibitem{huang2018}
G.~Huang, Z.~Liu, L.~van~der Maaten, and K.~Q. Weinberger,
\newblock ``Densely connected convolutional networks,'' 2018.

\bibitem{yu2019}
F.~Yu, D.~Wang, E.~Shelhamer, and T.~Darrell,
\newblock ``Deep layer aggregation,'' 2019.

\bibitem{Desplanques2020}
B.~Desplanques, J.~Thienpondt, and K.~Demuynck,
\newblock ``Ecapa-tdnn: Emphasized channel attention, propagation and
  aggregation in tdnn based speaker verification,''
\newblock {\em arXiv preprint arXiv:2005.07143}, 2020.

\bibitem{villalba22}
J.~Villalba, B.~J. Borgstr{\"o}m, S.~Kataria, M.~Rybicka, C.~D. Castillo,
  J.~Cho, L.~P. García-Perera, P.~A. Torres-Carrasquillo, and N.~Dehak,
\newblock ``{Advances in Cross-Lingual and Cross-Source Audio-Visual Speaker
  Recognition: The JHU-MIT System for NIST SRE21},''
\newblock in {\em Odyssey}, 2022, pp. 213--220.

\bibitem{Ronneberger2015}
O.~Ronneberger, P.~Fischer, and T.~Brox,
\newblock ``U-{N}et: Convolutional networks for biomedical image
  segmentation,''
\newblock in {\em Proc. MICCAI}, 2015, pp. 234--241.

\bibitem{Venkataramani2018}
S.~Venkataramani, J.~Casebeer, and P.~Smaragdis,
\newblock ``End-to-end source separation with adaptive front-ends,''
\newblock in {\em 2018 52nd Asilomar Conference on Signals, Systems, and
  Computers}. IEEE, 2018, pp. 684--688.

\bibitem{Quatieri2008}
T.~F. Quatieri,
\newblock {\em Discrete-Time Speech Signal Processing: Principles and
  Practice},
\newblock Pearson Education, 2008.

\bibitem{vctk}
C.~Valentini-Botinhao and J.~Yamagishi,
\newblock ``Speech enhancement of noisy and reverberant speech for
  text-to-speech,''
\newblock {\em IEEE/ACM Transactions on Audio, Speech, and Language
  Processing}, vol. 26, no. 8, pp. 1420--1433, 2018.

\bibitem{Ioffe2015}
S.~Ioffe and C.~Szegedy,
\newblock ``Batch normalization: Accelerating deep network training by reducing
  internal covariate shift,''
\newblock in {\em ICML}, 2015, pp. 448--456.

\bibitem{Hu2018}
J.~Hu, L.~Shen, and G.~Sun,
\newblock ``Squeeze-and-excitation networks,''
\newblock in {\em XCVPR}, 2018, pp. 7132--7141.

\bibitem{He2016}
K.~He, X.~Zhang, S.~Ren, and J.~Sun,
\newblock ``Deep residual learning for image recognition,''
\newblock in {\em Proceedings of the IEEE conference on computer vision and
  pattern recognition}, 2016, pp. 770--778.

\bibitem{librispeech}
V.~Panayotov, G.~Chen, D.~Povey, and S.~Khudanpur,
\newblock ``Librispeech: An asr corpus based on public domain audio books,''
\newblock in {\em ICASSP}, 2015, pp. 5206--5210.

\bibitem{timit}
J.~S. {Garofolo}, L.~F. {Lamel}, W.~M. {Fisher}, J.~G. {Fiscus}, and D.~S.
  {Pallett},
\newblock ``{DARPA TIMIT acoustic-phonetic continous speech corpus CD-ROM. NIST
  speech disc 1-1.1},'' Feb. 1993.

\bibitem{air}
M.~Jeub, M.~Schafer, and P.~Vary,
\newblock ``A binaural room impulse response database for the evaluation of
  dereverberation algorithms,''
\newblock in {\em International Conference on Digital Signal Processing}, 2009,
  pp. 1--5.

\bibitem{voice-home}
N.~Bertin, E.~Camberlein, E.~Vincent, R.~Lebarbenchon, S.~Peillon,
  E.~Lamand{\'e}, S.~Sivasankaran, F.~Bimbot, I.~Illina, A.~Tom, S.~Fleury, and
  E.~Jamet,
\newblock ``{A French corpus for distant-microphone speech processing in real
  homes},''
\newblock in {\em Interspeech}, 2016.

\bibitem{Voices}
C.~Richey, M.~A. Barrios, Z.~Armstrong, C.~Bartels, H.~Franco, M.~Graciarena,
  A.~Lawson, M.~K. Nandwana, A.~Stauffer, J.~van Hout, P.~Gamble, J.~Hetherly,
  C.~Stephenson, and K.~Ni,
\newblock ``Voices obscured in complex environmental settings (voices)
  corpus,'' 2018.

\bibitem{musan}
D.~Snyder, G.~Chen, and D.~Povey,
\newblock ``Musan: A music, speech, and noise corpus,'' 2015.

\bibitem{noisex}
A.~Varga and H.~J.~M. Steeneken,
\newblock ``Assessment for automatic speech recognition: Ii. noisex-92: A
  database and an experiment to study the effect of additive noise on speech
  recognition systems,''
\newblock {\em Speech Communication}, vol. 12, no. 3, pp. 247--251, 1993.

\bibitem{Jayant1974}
N.~S. Jayant,
\newblock ``Digital coding of speech waveforms: {PCM}, {DPCM}, and {DM}
  quantizers,''
\newblock {\em Proceedings of the IEEE}, vol. 62, no. 5, pp. 611--632, 1974.

\bibitem{wang2018}
D.-L. Wang and J.~Chen,
\newblock ``Supervised speech separation based on deep learning: An overview,''
\newblock {\em IEEE/ACM Transactions on Audio, Speech, and Language
  Processing}, vol. 26, no. 10, pp. 1702--1726, 2018.

\bibitem{Kolbaek2019}
M.~Kolbaek, Z.-H. Tan, and J.~Jensen,
\newblock ``On the relationship between short-time objective intelligibility
  and short-time spectral-amplitude mean-square error for speech enhancement,''
\newblock {\em IEEE/ACM Transactions on Audio, Speech, and Language
  Processing}, vol. 27, no. 2, pp. 283--295, 2019.

\bibitem{pesq}
A.~W. Rix, J.~G. Beerends, M.~P. Hollier, and A.~P. Hekstra,
\newblock ``Perceptual evaluation of speech quality ({PESQ})- {a} new method
  for speech quality assessment of telephone networks and codecs,''
\newblock in {\em ICASSP}, 2001, vol.~2, pp. 749--752.

\bibitem{taal2010}
C.~H. Taal, R.~C. Hendriks, R.~Heusdens, and J.~Jensen,
\newblock ``A short-time objective intelligibility measure for time-frequency
  weighted noisy speech,''
\newblock in {\em ICASSP}, 2010, pp. 4214--4217.

\bibitem{hu2007}
Y.~Hu and P.~C. Loizou,
\newblock ``Subjective comparison and evaluation of speech enhancement
  algorithms,''
\newblock {\em Speech Communication}, vol. 49, no. 7-8, pp. 588--601, 2007.

\bibitem{goodfellow2020}
I.~Goodfellow, J.~Pouget-Abadie, M.~Mirza, B.~Xu, David Warde-F., S.~Ozair,
  A.~Courville, and Y.~Bengio,
\newblock ``Generative adversarial networks,''
\newblock {\em Communications of the ACM}, vol. 63, no. 11, pp. 139--144, 2020.

\bibitem{speechbrain}
M.~Ravanelli, T.~Parcollet, P.~Plantinga, A.~Rouhe, S.~Cornell, L.~Lugosch,
  C.~Subakan, N.~Dawalatabad, A.~Heba, J.~Zhong, J.-C. Chou, S.-L. Yeh, S.-W.
  Fu, C.-F. Liao, E.~Rastorgueva, F.~Grondin, W.~Aris, H.~Na, Y.~Gao, R.~De
  Mori, and Y.~Bengio,
\newblock ``{SpeechBrain}: A general-purpose speech toolkit,'' 2021.

\bibitem{VoiceBank}
C.~Veaux, J.~Yamagishi, and S.~King,
\newblock ``The voice bank corpus: Design, collection and data analysis of a
  large regional accent speech database,''
\newblock in {\em International Conference Oriental COCOSDA}, 2013, pp. 1--4.

\bibitem{reddy2020}
C.~K.~A. Reddy, V.~Gopal, R.~Cutler, E.~Beyrami, R.~Cheng, H.~Dubey,
  S.~Matusevych, R.~Aichner, A.~Aazami, S.~Braun, et~al.,
\newblock ``The interspeech 2020 deep noise suppression challenge: Datasets,
  subjective testing framework, and challenge results,''
\newblock {\em arXiv preprint arXiv:2005.13981}, 2020.

\bibitem{hannun2014}
A.~Hannun, C.~Case, J.~Casper, B.~Catanzaro, G.~Diamos, E.~Elsen, R.~Prenger,
  S.~Satheesh, S.~Sengupta, and A.~Coates,
\newblock ``Deep speech: Scaling up end-to-end speech recognition,''
\newblock {\em arXiv preprint arXiv:1412.5567}, 2014.

\bibitem{whisper}
A.~Radford, J.~W. Kim, T.~Xu, G.~Brockman, C.~McLeavey, and I.~Sutskever,
\newblock ``Robust speech recognition via large-scale weak supervision,''
\newblock {\em arXiv preprint arXiv:2212.04356}, 2022.

\bibitem{sctk}
J.~Fiscus,
\newblock ``{SCTK}, the {NIST} scoring toolkit,''
\newblock 2021.

\end{thebibliography}
\end{document}